\documentclass[aps, pra, twocolumn, superscriptaddress]{revtex4-2}

\usepackage{graphicx}
\usepackage{amsmath}
\usepackage{amssymb}
\usepackage{nicefrac}
\usepackage{siunitx}
\usepackage[export]{adjustbox}
\usepackage{ragged2e}
\usepackage{multirow}
\usepackage{float}
\usepackage{xspace}
\usepackage{mathtools}
\usepackage{braket}
\usepackage{soul}

\usepackage[dvipsnames]{xcolor}
\usepackage{hyperref}
\hypersetup{
  colorlinks   = true,
  urlcolor     = MidnightBlue,
  linkcolor    = MidnightBlue,
  citecolor    = Maroon
}

\setcitestyle{numbers}

\renewcommand{\arraystretch}{2.0}

\begin{document}

\title{Modular fault-tolerant quantum computing on a non-CSS code}

\author{Robert Freund}
\altaffiliation{These authors contributed equally}
\affiliation{Universit\"{a}t Innsbruck, Institut f\"{u}r Experimentalphysik, Innsbruck, Austria}
\email[Email to ]{Robert.Freund@uibk.ac.at}

\author{Friederike Butt}
\altaffiliation{These authors contributed equally}
\affiliation{Institute for Theoretical Nanoelectronics (PGI-2), Forschungszentrum J\"{u}lich, J\"{u}lich, Germany}
\affiliation{Institute for Quantum Information, RWTH Aachen University, Aachen, Germany}

\author{César Benito}
\affiliation{Instituto de Fisica Teorica UAM-CSIC, Universidad Autonoma de Madrid, Cantoblanco, Madrid, Spain}

\author{Ivan Pogorelov}
\affiliation{Universit\"{a}t Innsbruck, Institut f\"{u}r Experimentalphysik, Innsbruck, Austria}

\author{Marcel Meyer}
\affiliation{Universit\"{a}t Innsbruck, Institut f\"{u}r Experimentalphysik, Innsbruck, Austria}

\author{Alex Steiner}
\affiliation{Universit\"{a}t Innsbruck, Institut f\"{u}r Experimentalphysik, Innsbruck, Austria}

\author{Alejandro Bermudez}
\affiliation{Instituto de Fisica Teorica UAM-CSIC, Universidad Autonoma de Madrid, Cantoblanco, Madrid, Spain}

\author{Markus M\"{u}ller}
\affiliation{Institute for Theoretical Nanoelectronics (PGI-2), Forschungszentrum J\"{u}lich, J\"{u}lich, Germany}\affiliation{Institute for Quantum Information, RWTH Aachen University, Aachen, Germany}

\author{Thomas Monz}
\affiliation{Universit\"{a}t Innsbruck, Institut f\"{u}r Experimentalphysik, Innsbruck, Austria}
\affiliation{Alpine Quantum Technologies GmbH, Innsbruck, Austria}

\date{\today}

\begin{abstract}
Modularization promises to break down the design and implementation complexity of large scale quantum processors into smaller manageable subtasks.
In this approach, quantum channels, realized for instance through physical rerouting of qubits or quantum teleportation, connect multiple modules.
Each of those modules hosts a subset of qubits, e.g. multiple logical qubits, and provides quantum operations on them.
In this work, we implement for the first time all logical operations required for modular fault-tolerant universal quantum computing with a non-Calderbank-Shor-Steane (CSS) code, the perfect 
$[[5, 1, 3]]$ code, on a trapped-ion quantum computer.
This code is the smallest quantum error-correcting (QEC) code capable of correcting any single-qubit error, making it a compact alternative to larger CSS codes, provided that the remaining logical primitives are available.
We demonstrate logical state teleportation and a full suite of fault-tolerant operations required for universal logical control, including logical state preparation, QEC with real-time feedback, logical measurements, magic-state preparation, logical entangling operations, and magic-state injection. Moreover, we characterize the logical spectator error picked up by idling logical qubits during quantum operations on distinct qubit registers and demonstrate a logical Pauli quantum process tomography that minimizes the required sampling resources for logical tomography.

\end{abstract}

\maketitle

\section{Introduction}

\begin{figure*}[!tb]
    \centering
    \includegraphics[width=170mm]{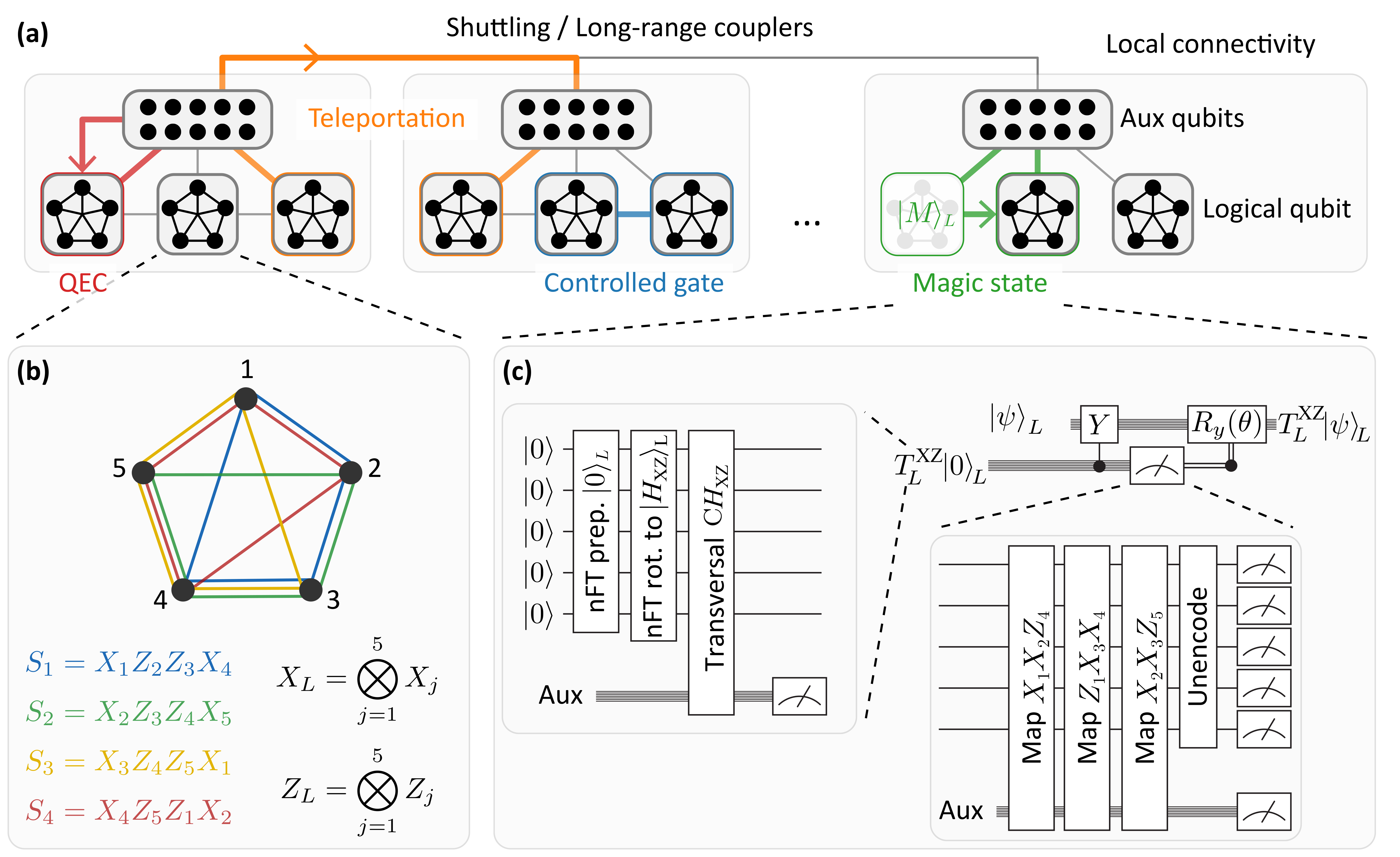}
    \caption{\justifying \textbf{Modular quantum computing architecture using the $[[5, 1, 3]]$ code.} (a) Scalable quantum computing architecture based on a modular design featuring smaller units with local connectivity within the logical qubit and ancilla qubit registers. The ancilla qubits of different units are connected via long-range couplers or shuttling. (b) The four weight-4 stabilizers of the $[[5, 1, 3]]$ code are cyclic permutations of $S^{(1)} = X_1 Z_2 Z_3 X_4 I_5$. The logical Pauli operations can be implemented by applying $X$ and $Z$ to all five physical qubits. (c) Logical operations are decomposed into physical operations for the example of magic state injection. This operation realizes a logical $\pi/4$-rotation around the $Y$ axis labeled $T^{XZ}_\mathrm{L}$. FT magic state preparation (left) consists of the non-FT magic state preparation and a verification by means of a logical measurement. The FT measurements (right) are split into mapping the logical operator $-Z_L$ three times and unencoding the logical qubit which is necessary due to the non-CSS properties of the error correction code~\cite{ryan2022implementing}.
    }
    \label{fig:overview}
\end{figure*}

Quantum error correction (QEC) is a central requirement for large-scale and reliable quantum computation. QEC enables the detection and correction of local physical errors while preserving the logical information by encoding it into non-local degrees of freedom. Recent experiments have demonstrated increasingly long-lived logical memories and fault-tolerant primitives~\cite{google2025quantum, lacroix2025scaling, caune2024demonstrating, putterman2024hardware, butt2026demonstration, postler2022demonstration, reichardt2024demonstration, bluvstein2024logical, reichardt2024fault, bluvstein2025fault, mathiot2026benchmarking, lin2026surface, besedin2026lattice, wang2026superconducting}. However, scalable fault-tolerant (FT) quantum computation requires not only efficient encoding and protecting stored logical information but also rerouting, connecting, and manipulating encoded states across distinct logical registers, calling for a modular approach.

Logical state teleportation is a method for entangling potentially distant modules within a larger architecture without physically shuttling or swapping full registers of physical qubits. Instead, quantum information is transferred through logical measurements and classically controlled Pauli corrections~\cite{gottesman1999demonstrating, Nielsen_and_Chuang} without the need to directly couple source and target registers. Logical teleportation is a key operation in modular hardware architectures with separated code blocks or operational zones, as it only requires a shared entanglement resource or the coupling to auxiliary qubits~\cite{monroe2014large, yoder2025tour}. Figure~\ref{fig:overview}(a) depicts a modular scheme featuring local connectivity between logical qubit blocks and auxiliary registers. The modules are interconnected through coupling the respective auxiliary registers with one another.
Many recent proposals for large-scale architectures rely on these modular design principles in combination with Calderbank--Shor--Steane (CSS) quantum low-density parity check (qLDPC) codes~\cite{yoder2025tour, webster2025explicit, webster2026pinnacle, cain2026shor}. Modularity allows the scaling of quantum computation to a large number of logical qubits~\cite{kielpinski2002architecture} while qLDPC codes feature high encoding rates of logical qubits into physical qubits. 
Logical teleportation based on topological CSS codes has recently been demonstrated on superconducting, trapped-ion and neutral-atom platforms~\cite{bluvstein2025fault, lacroix2025scaling, ryan2024high, lin2026surface, besedin2026lattice, wang2026superconducting}. Here, the code structure inherently simplifies the FT implementation of logical measurements and lattice-surgery-style~\cite{horsman2012surface} operations.
Extending such protocols to less explored code families, such as non-CSS and qLDPC codes, remains to be demonstrated and can provide new routes towards resource efficient logical quantum processing.

In this work, we fill in this gap by demonstrating FT modular logical state teleportation between two non-CSS code blocks on a trapped-ion quantum processor, specifically two perfect 5-qubit code blocks. The protocol realizes a modular primitive for moving quantum information at the logical level, while avoiding a direct physical relocation or coupling of encoded registers. 
The required logical measurements and classically controlled feedback operations are applied in real-time.
Beyond logical state teleportation, we demonstrate FT logical state preparation, QEC on multiple logical qubits, FT logical measurements, as well as new protocols for FT magic-state preparation, FT entangling logical operations and magic-state injection on the 5-qubit code. 
We also introduce and demonstrate a logical Pauli quantum process tomography to characterize the logical primitives that allows to separate state preparation and readout errors from those of the logical operations with minimal resource requirements. Our work constitutes the first experimental implementation of modular universal logical quantum computation for a non-CSS code.

This paper is structured as follows. First, we introduce the experimental trapped-ion platform in Sec.~\ref{sec:experimental_platform}. In Sec.~\ref{sec:toolbox}, we summarize all implemented protocols, discuss their performance, and compare the results to numerical simulations. Lastly, in Sec.~\ref{sec:characterization}, we characterize the demonstrated building blocks by extracting state preparation and measurement (SPAM) error free performance metrics for logical qubits before concluding in Sec.~\ref{sec:dicsussion}.

\section{Experimental Setup}\label{sec:experimental_platform}

The experiment is performed on a 16-qubit trapped ion quantum computing device. 
A chain of 16 $^{40}$\textrm{Ca}$^+$ ions is confined in a linear Paul trap.
The physical qubits are encoded in the $\ket{0}=\ket{4 ^2\textrm{S}_{1/2}, m_J = -1/2}$ and $\ket{1}=\ket{3 ^2\textrm{D}_{5/2}, m_J = -1/2}$ Zeeman sub-levels. 
Quantum gates are realized by optically addressing individual ions with laser light. On resonance with the qubit transition, timed Rabi oscillations implement single qubit gates. The M{\o}lmer-S{\o}rensen (MS) interaction~\cite{sorensen2000entanglement} provides two-qubit gates between any desired ion pair in the chain. Therefore, the native gate set of the device includes arbitrary-angle rotation gates $R(\theta,\phi) = \exp(-i\frac{\theta}{2}[X \cos \phi + Y \sin \phi])$, $Z$-gates \hbox{$R_Z(\theta)=\exp(-i\frac{\theta}{2}Z)$}, and maximally-entangling two-qubit gates $XX(\pi/2) = \exp(- i\frac{\pi}{4} X \otimes X)$ with $X$, $Y$ and $Z$ being Pauli operators. $Z$-gates are carried out virtually by updating the phase of the light fields implementing subsequent quantum gates. All circuits shown are decomposed into this native gate set and whenever possible quantum gates are contracted to shorten the circuits. A more detailed description of the apparatus is given in Refs.~\citenum{pogorelov2021compact, postler2022demonstration, heussen2023strategies}.

Furthermore, the setup provides mid-circuit measurements and outcome-conditional feed-forward operations, which are crucial for many fault-tolerant quantum computing paradigms. 
The mid-circuit measurement, processing, and conditional circuit adaption are executed within a single circuit execution, thus in real-time. 
In this work, we bundle multiple mid-circuit measurements into a single mid-circuit measurement operation to reduce time overheads and adhere to implementation restrictions (details are provided in Ref.~\citenum{postler2024demonstration}). Due to this bundling, the qubit registers are split into auxiliary (measured) and data (not measured) qubits throughout this work as further outlined in App.~\ref{app:Mid-circuit_process}.

\section{Toolbox for the $[[5, 1, 3]]$ code}
\label{sec:toolbox}

The $[[5, 1, 3]]$ code~\cite{laflamme1996perfect} is a non-CSS stabilizer code that encodes $k=1$ logical qubit in $n=5$ physical qubits at a code distance of $d=3$, such that any single error can be corrected. 
CSS codes can be constructed by combining two classical linear codes where each classical code provides the underlying $X$- and $Z$-stabilizer structure of the resulting quantum code~\cite{calderbank1996good, steane1996multiple}. As a consequence, $Z$- and $X$-errors can be corrected independently and CNOT gates are inherently transversal~\cite{gottesman1997stabilizer}. In contrast, non-CSS codes generally do not support transversal CNOT gates or independent $X$- and $Z$-error-correction, but they can offer more compact encodings. For example, the smallest error-correcting CSS stabilizer code is the $[[7, 1, 3]]$ code~\cite{steane1997active}, while the same distance is achieved with five qubits in the non-CSS $[[5,1,3]]$ code, fulfilling the quantum singleton bound~\cite{knill1997theory}. 
The stabilizers are
\begin{align}
    S_1 &= X_1 Z_2 Z_3 X_4 I_5, \label{eq:stabilizers}\quad 
    S_2 = I_1 X_2 Z_3 Z_4 X_5, \\
    S_3 &= X_1 I_2 X_3 Z_4 Z_5, \quad
    S_4 = Z_1 X_2 I_3 X_4 Z_5\nonumber, 
\end{align}
and logical operators $X_{\mathrm{L}} = \bigotimes_{j=1}^{5}  X_j$ and $Z_{\mathrm{L}} = \bigotimes_{j=1}^{5} Z_j$, as illustrated in Fig.~\ref{fig:overview}(b). Stabilizer-equivalent representations of the logical operators are of weight 3 and consist of both physical $X$ and $Z$ operations, for example $\Tilde{X}_{\mathrm{L}} = Y_2 Y_3 X_5$. 
The $[[5,1,3]]$ code admits fault-tolerant implementations of the full single-qubit Clifford group using transversal single-qubit gates together with fixed qubit-permutations that are carried out in software. In our qubit ordering, the logical Hadamard gate $H_\mathrm{L}$ and phase gate $S_\mathrm{L}$ can be implemented by applying a transversal operation $H^{\otimes 5}$ and $(S^\dagger)^{\otimes 5}$, respectively, followed by the permutation $(1, 2, 3, 4, 5) \longrightarrow (2,5,3,1,4)$~\cite{abobeih2022fault}. Applying the indicated permutation $(1, 2, 3, 4, 5) \longrightarrow (2,5,3,1,4)$ twice preserves the original stabilizer group, generated by the operators specified in \eqref{eq:stabilizers}. Therefore, the product $H_\mathrm{L}S_\mathrm{L}$ is implemented by
\begin{equation}
    H_\mathrm{L}S_\mathrm{L} = H^{\otimes 5}(S^\dagger)^{\otimes 5},
\end{equation}
without an additional qubit permutation~\cite{yoder2016universal}. 

In the following, we present a toolbox for FT universal quantum computing with the $[[5, 1, 3]]$ code and decompose logical into physical operations as exemplified in Fig.~\ref{fig:overview}(c). This includes all building blocks required for implementing a quantum memory, state teleportation, as well as magic state injection. 

\subsection{Quantum memory}

First, we present the FT protocols for logical Pauli-state preparation, QEC and logical measurements on the $[[5, 1, 3]]$ code and demonstrate them on a trapped-ion quantum processor. These three building blocks have already been demonstrated independently from this work on a trapped-ion device~\cite{ryan2022implementing}. 

\subsection*{Fault-tolerant Pauli-state preparation} 
Logical Pauli states can be initialized by means of a non-FT initialization followed by a verification step. This verification consists of a measurement of suitable logical operators. It flags the possible presence of dangerous errors,~i.e., errors that cause a logical failure if not detected. If a potentially dangerous error is signaled by a raised flag, the run is discarded and a new logical state is prepared. We use known constructions for FT preparation of logical Pauli eigenstates~\cite{ryan2022implementing, zen2024quantum}. 
Specifically, a logical qubit is prepared in $\ket{1}_\mathrm{L}$ or $\ket{-}_\mathrm{L}$ using the FT circuits with only two physical auxiliary qubits as shown in Fig.~\ref{fig:ft_init} of App.~\ref{app:detailed_FT_teleportation_protocol}. The states are rotated fault-tolerantly to $\ket{0}_\mathrm{L}$ and $\ket{+}_\mathrm{L}$ by applying transversal operations $X_\mathrm{L}$ or $Z_\mathrm{L}$ gate, respectively. The states $\ket{+i}_\mathrm{L}$ and $\ket{-i}_\mathrm{L}$ are generated by initializing $\ket{0}_\mathrm{L}$ and applying logical $H_\mathrm{L}$ and $S_\mathrm{L}$ gates, which do not change the weight of a pre-existing single error.

\subsection*{Fault-tolerant QEC}
\begin{figure}[!tb]
    \centering
    \includegraphics[width=90mm]{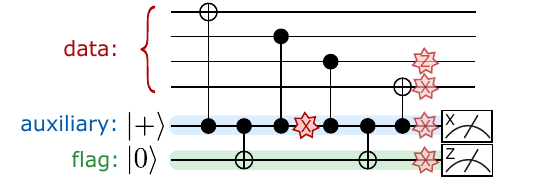}
    \caption{\justifying \textbf{Circuit for fault-tolerant stabilizer measurement~\cite{ryan2022implementing, reichardt2020fault}.} An $XZZX$ stabilizer is measured by coupling an auxiliary qubit to the data qubits that support the respective operator. An additional flag qubit is coupled to the first auxiliary qubit such that potentially dangerous errors (red) also propagate onto the flag qubit and become detectable. }
    \label{fig:Flag_QEC_circuit}
\end{figure}

The $XZZX$-type stabilizers of the $[[5, 1, 3]]$ code can be measured fault-tolerantly using flagged circuits as shown in Fig.~\ref{fig:Flag_QEC_circuit}~\cite{reichardt2020fault, ryan2022implementing}. Here, an additional flag qubit heralds the presence of a potentially dangerous propagated error that otherwise, if not detected and corrected, would lead to a failure.
As a more compact alternative, we use a single flag qubit to measure a pair of stabilizers, which has been shown to be FT for a specific gate ordering~\cite{reichardt2020fault}. The corresponding circuit is shown in Fig.~\ref{fig:stabilizer_measurement} of App.~\ref{app:FlagQEC}. If no flag is raised and the syndrome is trivial, no correction is applied. For non-trivial syndromes and no risen flags, the stabilizers are remeasured without flags and the lookup table (LUT) in Tab.~\ref{tab:LUT_QEC_standard} of App.~\ref{app:LookupTables} is used. If a flag is raised, the syndrome is remeasured and a flag-based LUT, Tab.~\ref{tab:LUT_QEC_parallel} of App.~\ref{app:LookupTables}, is applied, falling back to the standard LUT if no match is found. Stabilizer measurements on the $[[5, 1, 3]]$ code have been implemented on a trapped-ion  quantum processor~\cite{ryan2022implementing} and in diamond NV centers~\cite{abobeih2022fault}. 

\begin{figure*}[!t]
    \centering
    \includegraphics[width=180mm]{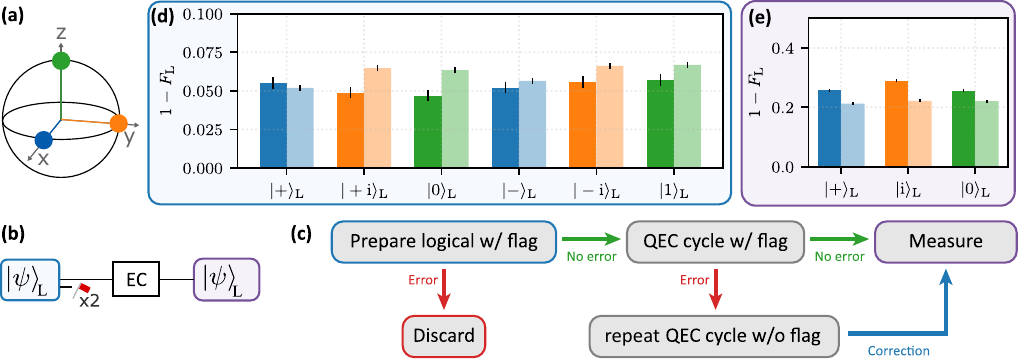}
    \caption{\justifying \textbf{Fault-tolerant state preparation and quantum memory.} (a) $\pm1$ eigenstates of the three logical Pauli operators $X, Y$ and $Z$ in the Bloch-sphere representation. (b) A Pauli eigenstates is prepared and followed by a round of fault-tolerant QEC and a logical measurement, together realizing a quantum memory. (c) The decision tree outlining the mid-circuit classical logic of the FT QEC cycle on a single logical qubit including potential real-time corrections. (d) Experimental logical state infidelities (dark) for the six cardinal states after FT logical state preparation and a FT logical measurement in the respective basis. Results from numerical simulations (light) are shown alongside for comparison. The error bars are given by $1\sigma$ Clopper-Pearson confidence intervals. (e) Logical state infidelities after FT initialization, a single round of FT QEC and a FT measurement.}
    \label{fig:single_qubit_qec}
\end{figure*}

\begin{figure}[!tb]
    \centering
    \includegraphics[width=85mm]{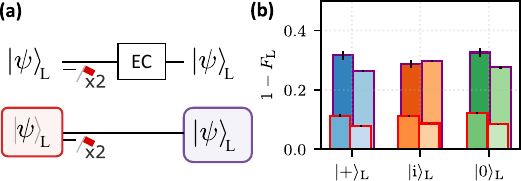}
    \caption{\justifying \textbf{Logical spectator error.} (a) We prepare two logical qubits and measure the infidelity of the lower qubit before (red box) and after (purple box) performing a QEC cycle on the upper qubit. (b) Infidelities of the lower logical qubit before (red-outlined bars in the front) and after (bars in the background) the QEC cycle on the upper logical qubit  is applied. The uncertainty of the infidelity is given by $1\sigma$ Clopper-Pearson confidence intervals.}
    \label{fig:results_crosstal}
\end{figure}

\subsection*{Fault-tolerant logical measurement}

For CSS codes, transversal measurements allow the simultaneous extraction of stabilizers and logical operators from a single destructive readout. However, the $[[5,1,3]]$ code is non-CSS, and we have to use a dedicated fault-tolerant scheme to reliably extract logical observables~\cite{ryan2022implementing}. 

Specifically, for the $[[5,1,3]]$ code, we map three distinct representations of the respective logical operators onto auxiliary qubits with flags, followed by unencoding and a destructive measurement of the data qubits. In doing so, we obtain three logical measurement outcomes from the mappings to the auxiliary qubits as well as the full syndrome and another logical measurement outcome from the measurement of the data qubits. If the first three logical measurement outcomes obtained from the measured auxiliary qubits agree with each other and no flag is raised, their value is accepted; otherwise, the final logical outcome is corrected based on the inferred syndrome. In contrast to previous works~\cite{ryan2022implementing}, we use a single flag qubit instead of two to measure all three representations fault-tolerantly using the circuits shown in Fig.~\ref{fig:circuit_measure_logicals} of App.~\ref{app:detailed_FT_teleportation_protocol}.

We combine the FT preparation of a single logical qubit in a Pauli eigenstate (shown in Fig.~\ref{fig:single_qubit_qec}(a)) with the FT logical measurement in the respective Pauli basis to extract logical state preparation and measurement (SPAM) error. In Fig.~\ref{fig:single_qubit_qec}(d), the logical state fidelities for six different input states measured in the experiment (dark) are shown. We postselect on non-trivial flags during the verification of the logical input state, which leads to an averaged acceptance rate of 80\,\% and an average infidelity of $5.24^{+0.16}_{-0.14} \,\%$. Experiments are accompanied by a numerical simulation (light) using an experimentally informed error model as summarized in App.~\ref{app:numerical_methods}. The simulation predicts an average infidelity of $6.16 ^{+0.08}_{-0.08} \,\%$ with an average acceptance rate of 77\,\%. Here, and in the remainder of this work, the uncertainty in simulations and experiment are given by Clopper-Pearson confidence interval assuming a binomial distribution of the measurement outcomes with the accepted number of shots and the outcome probability~\cite{Clopper1934Uncertainty}. We report $1\sigma$-equivalent confidence intervals. Values derived from binomial distributed measurement results are assigned an uncertainty, which is derived using standard error propagation unless stated otherwise. For all reported numerical fidelities in this work we provide the total number of accepted shots of the underlying measurements in App.~Tab.~\ref{tab:post-selected-total-shots}. These logical SPAM errors of a single logical qubit serve as a infidelity baseline in further experiments, e.g., the quantum memory experiment.

\subsection*{Implementation of a quantum memory}

We probe error-correction capabilities by extending the sequence of FT preparation and FT measurement by an inserted round of QEC as shown in Fig.~\ref{fig:single_qubit_qec}(b) and (c). The state infidelity after one round of QEC is shown in Fig.~\ref{fig:single_qubit_qec}(e) for different logical Pauli eigenstates and averages to $26.8^{+0.3}_{-0.3} \, \% $ (simulations: $21.9^{+0.3}_{-0.3} \, \%)$ with an acceptance rate of 75\,\% (77\,\%). As we only postselect on the logical state preparation but not on the measured error syndromes, the acceptance rate is comparable to the previous experiment. The fidelity is mostly limited by two-qubit gate errors as described in App.~\ref{app:ErrorBudget}. 

\begin{figure*}[!tb]
    \centering
    \includegraphics[width=180mm]{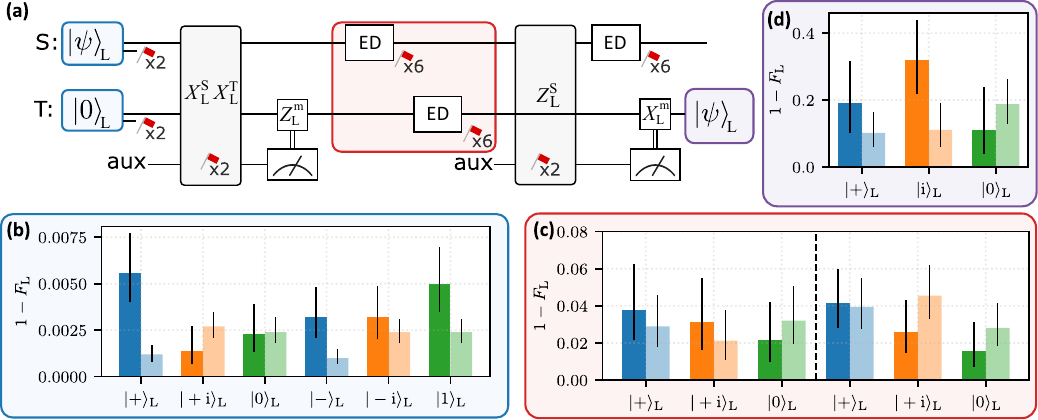}
    \caption{\justifying \textbf{Fault-tolerant logical state teleportation. }(a) A schematic circuit for FT logical state teleportation from the source (S) to target (T) qubit is drawn. Logical input states $|\psi\rangle_{\mathrm{L}} = |+\rangle_{\mathrm{L}}, |+i\rangle_{\mathrm{L}}, |0\rangle_{\mathrm{L}}$ (blue, orange, green) are prepared on the source register. Logical state fidelities at three different stages during the circuit are presented, corresponding to the blue, red and purple box. (b) The logical state infidelities from experiment (dark) and simulations (light) for FT logical state preparation and measurement of a single logical qubit are shown. (c) FT error detection (ED) with flags on both logical qubits is performed after FT preparation and the characterization concludes with a FT logical measurement of a single logical qubit. The infidelities are shown with the source qubit on the left and target qubit on the right. (d) The logical state infidelity of the teleported state is presented. The uncertainty in the three plots is given by $1\sigma$ Clopper-Pearson confidence interval.}
    \label{fig:results_teleportation}
\end{figure*}

\subsection*{Logical spectator error}

Extending the system to two logical qubits allows us to quantify logical crosstalk for the first time in the 5-qubit code. We explicitly measure how one logical qubit gets affected while FT circuits are run on the other one. The FT building blocks introduced above are composed to prepare two logical qubits in the Pauli eigenbasis. Then, one cycle of FT QEC is applied on the first logical qubit after which the second logical qubit is measured, as shown in Fig.~\ref{fig:results_crosstal}(a). The state infidelity of the second logical qubit directly after its initialization is depicted in Fig.~\ref{fig:results_crosstal}(b) (red framed bars in the front) with experimental values on the left (dark) and simulated values on the right (light). Performing one FT QEC cycle increases the state infidelity of the spectator logical qubit by roughly 20\,\% as shown by the bars in the background (purple frame). The discrepancy is explained by error sources such as dephasing of idling qubits or mid-circuit-measurement-induced errors. Note that all gate operations are executed sequentially and parallel execution could substantially reduce idling times. Appendix~\ref{app:ErrorBudget} provides additional details on the error contributions.

\subsection{Fault-tolerant teleportation} 

Having established logical single-qubit Clifford control, we now describe the FT state teleportation protocol which can connect two modules in a modular quantum computing architecture. State teleportation is used to transfer a logical state from a source (S) to a target (T) qubit. The teleportation protocol consists of two logical measurements~\cite{Nielsen_and_Chuang, gottesman1999demonstrating}: first the joint operator $X_\mathrm{L}^\mathrm{(S)} X_\mathrm{L}^\mathrm{(T)}$ is measured, followed by a measurement of the source logical operator $Z_\mathrm{L}^\mathrm{(S)}$ (see Fig.~\ref{fig:results_teleportation}(a)). 
These measurements randomly project the state into either a $+1$ or $-1$ eigenstates of the respective logical operators. 
The corresponding outcomes determine if $Z_\mathrm{L}^\mathrm{(T)}$ or $X_\mathrm{L}^\mathrm{(T)}$ operators need to be applied to the target register to complete the teleportation. We need to account for three classes of single faults to achieve both logical measurements fault-tolerantly. 

First, a fault in the measurement of either $X_\mathrm{L}^\mathrm{S} X_\mathrm{L}^\mathrm{T}$ or $Z_\mathrm{L}^\mathrm{S}$ can flip the logical measurement outcome, causing a logical error. We thus measure each logical operator at least twice (App.~\ref{app:detailed_FT_teleportation_protocol}, Fig.~\ref{fig:teleportation_detailed}). If the two measured values are not the same, we run one round of QEC on those registers that participate in the respective logical measurement, and re-measure the logical operators (App.~\ref{app:detailed_FT_teleportation_protocol}, Fig.~\ref{fig:protocol}(a)). 

Second, a fault on an auxiliary qubit used for a logical measurement can propagate to multiple data qubits and produce an uncorrectable error. We therefore use flagged logical measurement circuits. In combination with subsequent stabilizer measurements, we can prevent the undetected propagation of dangerous errors compromising FT. 

Third, during the mapping of $X_\mathrm{L}^\mathrm{(S)} X_\mathrm{L}^\mathrm{(T)}$ or $Z_\mathrm{L}^\mathrm{(S)}$ to the auxiliary register, a single error on one of the data qubits can propagate to an auxiliary qubit and flip the measured logical outcome. These errors are corrected by an intermediate round of QEC on the source and target qubit after the repeated logical measurements as depicted in Fig.~\ref{fig:protocol}(a) of App.~\ref{app:detailed_FT_teleportation_protocol}. The full protocol, lookup tables, and circuits are detailed in App.~\ref{app:detailed_FT_teleportation_protocol}. 

In the following, we deal with errors using postselection to maintain technical feasibility for the complex decision tree of deep logical protocols. The fault-tolerant stabilizer readouts, logical measurements and mapping of logical operators are accepted only if all syndromes and flags are trivial. The full structure of the protocol is illustrated in Fig.~\ref{fig:protocol}(a) of App.~\ref{app:detailed_FT_teleportation_protocol} while the number of decisions using postselection is much lower as shown in Fig.~\ref{fig:protocol}(b) of App.~\ref{app:detailed_FT_teleportation_protocol}. The protocol using postselection is shown schematically in Fig.~\ref{fig:results_teleportation}(a). Crucially, the Pauli corrections of the teleportation are executed with real-time feedback. Before we report on the implementation of the full teleportation protocol, we characterize its core building blocks. The logical state infidelity for FT preparation and measurement of one logical qubit is shown in Fig.~\ref{fig:results_teleportation}(b) and averages to $0.35^{+0.07}_{-0.05} \, \%$ (simulations: $0.20^{+0.03}_{-0.02} \, \%$) with an average acceptance of 46\,\% (39\,\%). These results provide a lower infidelity and acceptance rate compared to the previous results in Fig.~\ref{fig:single_qubit_qec}(d) because we use error detection instead of error correction. The characterization serves as a logical SPAM-error estimate and a baseline in infidelity for, e.g., the logical state teleportation.

As a next step, we prepare two logical qubits and perform one round of quantum error detection (QED) on each of them, followed by a FT measurement of one logical qubit. The corresponding logical state fidelities are shown in Fig.~\ref{fig:results_teleportation}(c). The results for the two logical qubits are separated by a dashed line. The average logical infidelity is $2.9^{+0.8}_{-0.5} \, \%$ (simulations: $3.3^{+0.7}_{-0.5} \, \%$) with an acceptance rate of 0.85\,\% (1.15\,\%). The infidelity increases compared to the SPAM error due to additional mid-circuit measurements and dephasing during idling.

The logical state fidelity after the logical state teleportation is shown in Fig.~\ref{fig:results_teleportation}(d) for three different input states. We find an average logical infidelity of $21^{+7}_{-5}\,\%$ (simulations: $13^{+4}_{-3}\,\%$) and an acceptance rate of 0.035\,\% (0.044\,\%). The infidelity of the teleportation is mainly limited by mid-circuit measurement errors as shown in App.~\ref{app:ErrorBudget}.

\subsection{Fault-tolerant non-Clifford gates}
\begin{figure*}[!tb]
    \centering
    \includegraphics[width=180mm]{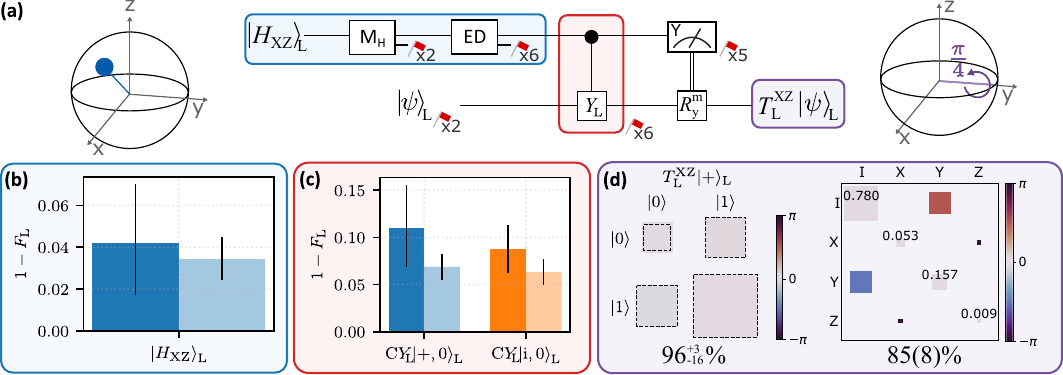}
    \caption{\justifying \textbf{Fault-tolerant magic state preparation and injection. }(a) The schematic shows the FT preparation of a magic state $|H_{\mathrm{XZ}}\rangle_{\mathrm{L}}$ in the blue box and the corresponding Bloch-sphere representation on the left. The $T^{\mathrm{XZ}}_{\mathrm{L}}$ gate is applied to an arbitrary input state $|\psi\rangle_\mathrm{L}$ by magic state injection. The equivalent rotation on the Bloch sphere is shown on the right. (b) Experimental (dark) and numerically simulated (light) logical state infidelity of a fault-tolerantly prepared magic state $|H_{\mathrm{XZ}}\rangle_{\mathrm{L}} = \cos{(\pi/8)} |0\rangle_{\mathrm{L}} + \sin{(\pi/8)|1\rangle}_{\mathrm{L}}$ is presented. (c) A logical pieceably FT C$Y_\mathrm{L}$ gate is characterized by the state infidelities of the entangled state for logical input states $|+, 0\rangle_{\mathrm{L}}$ and $|+i, 0\rangle_{\mathrm{L}}$. In both figures, the uncertainty is given by $1\sigma$ Clopper-Pearson confidence intervals. (d) The logical C$Y_{\mathrm{L}}$ gate is used to implement the logical $T^{\mathrm{XZ}}_{\mathrm{L}}$ gate. We perform logical state tomography for the prepared state $T^{\mathrm{XZ}}_{\mathrm{L}}\ket{+}_{\mathrm{L}}$ determining the density matrix on the left and reaching a state fidelity of $96^{+3}_{-16} \, \%$. The black dashed boxes correspond to the fault-free ideal outcomes. On the right, the noise process matrix of the $T^{\mathrm{XZ}}_{\mathrm{L}}$ gate determined in Sec~\ref{sec:characterization} is shown and a process fidelity of $85 \pm 8 \, \%$ is calculated. Since the values of the matrices are complex numbers, the absolute value determines the size of the square and the color represents the phase.
    }
    \label{fig:results_magic_states}
\end{figure*}

Next we complete the universal gate set by presenting the FT protocols for logical magic state preparation and an entangling gate between two logical qubits. The two building blocks are combined for magic-state injection realizing a non-Clifford gate.

\subsection*{Fault-tolerant magic state preparation}
We complete a FT universal set of gates by means of magic-state injection~\cite{bravyi2005universal, gottesman1999demonstrating}, which requires the FT preparation of specific resource states. We prepare the logical magic state 
\begin{align}
    |H_{\mathrm{XZ}}\rangle_\mathrm{L} = \cos{\left({\frac{\pi}{8}}\right)} |0\rangle_\mathrm{L} + \sin{\left({\frac{\pi}{8}}\right)} |1\rangle_\mathrm{L},
\end{align}
which is the $+1$ eigenstate of the logical Hadamard operator. We construct new circuits for the preparation of this magic state involving a verification and postselection on non-trivial stabilizer measurement outcomes, similar to previous experimental demonstrations for CSS codes~\cite{lacroix2025scaling, rosenfeld2025magic, postler2022demonstration}. We first prepare $|H_{XZ}\rangle_\mathrm{L}$ using a novel non-FT circuit. Then, we perform a FT measurement of $H_\mathrm{L}$ using a flagged circuit, and finally run one round of error detection. The scheme is illustrated in Fig.~\ref{fig:results_magic_states}(a) (blue box) and the full circuit is shown in Fig.~\ref{fig:nft_h} of App.~\ref{app:MS_injection_details}. Here, the error-detection step consists of one application of the parallel syndrome-extraction circuit shown in Fig.~\ref{fig:stabilizer_measurement} of App.~\ref{app:detailed_FT_teleportation_protocol}, followed by postselection on trivial syndrome outcomes and the absence of raised flags. 
We theoretically verify fault tolerance of the complete preparation circuit by inserting all possible single faults and checking that no single fault can lead to a logical error.

We determine a logical state infidelity of the prepared magic state of $4^{+3}_{-2} \, \%$ (simulations: $3.4^{+1.0}_{-1.0} \, \%$) with an average acceptance rate of 2.3 (7.3)\,\%, as shown in Fig.~\ref{fig:results_magic_states}(b). The results agree with the simulations within one standard deviation and serve as a baseline for the realization of a non-Clifford gate via magic state injection.

\subsection*{Pieceably fault-tolerant entangling gate}
The $[[5, 1, 3]]$ code is not a CSS code and therefore does not support a transversal logical two-qubit entangling gate. We instead implement a logical entangling gate by means of a \emph{pieceably} FT construction~\cite{yoder2016universal}. The central idea in pieceable fault tolerance is to divide a non-FT circuit into smaller pieces, each of which is FT up to correctable errors, and to insert QEC between these pieces. 

We implement a logical gate C$Z_\mathrm{L}$ using a round-robin construction~\cite{ryan2022implementing, yoder2016universal} as shown in Fig.~\ref{fig:ft_cz} of App.~\ref{app:MS_injection_details}. 
A round-robin gate on a distance-$d$ code consists of $d$ layers of transversal gates that act on qubits supporting a representation of the respective logical operator, such that the stabilizer group is preserved in total by construction. 
Specifically, we apply local basis changes to a subset of physical qubits for convenience. Then, we apply three layers of C$Z$ gates between two $[[5, 1, 3]]$ codes, and, after the second layer, we insert an intermediate error-detection block. Appendix~\ref{app:MS_injection_details} summarizes details on the implementation of the intermediate round of stabilizer extraction. 

The C$Z_\mathrm{L}$ gate is converted to a C$Y_\mathrm{L}$ gate by applying a transversal $S_\mathrm{L}^\dagger H_\mathrm{L}$ gate to the target qubit before and a $H_\mathrm{L}S_\mathrm{L}$ gate to the target qubit after the entangling operation. The C$Y_\mathrm{L}$ gate transforms the initial states $\ket{+}_{\mathrm{L}}\ket{0}_{\mathrm{L}}$ and $\ket{+i}_{\mathrm{L}}\ket{0}_{\mathrm{L}}$ into maximally entangled states and both logical qubits are measured subsequently. Figure~\ref{fig:results_magic_states}(c) shows the logical state fidelity of the maximally entangled logical states averaging to $10^{+3}_{-2} \, \%$ (simulations: $6.6^{+1.0}_{-1.0} \, \%$) and we find an acceptance rate of 0.36 (0.67)\,\%. The fidelity is comparable to the simulation and mainly limited by mid-circuit measurement errors as described in App.~\ref{app:ErrorBudget}.

\subsection*{Fault-tolerant magic-state injection}

We combine the ingredients above to implement a logical non-Clifford gate by means of magic-state injection, as illustrated in Fig.~\ref{fig:results_magic_states}(a). 
The target operation is
\begin{equation}
    T^{\mathrm{XZ}}_\mathrm{L} = e^{-i\pi Y_\mathrm{L}/8}.
\end{equation}
We first prepare a verified logical magic state on an auxiliary code block. The auxiliary block is then entangled with a data-qubit register using the above logical C$Y_\mathrm{L}$ operation. Finally, the auxiliary block is measured in the logical $Y$ basis. Depending on the measurement outcome, a logical $R_Y$ rotation with angle $\pi/2$ is applied to the data block.

We perform logical state tomography after the full magic-state injection protocol for the input state $|+\rangle_\mathrm{L}$, as shown in Fig.~\ref{fig:results_magic_states}(d), and find a logical state infidelity of $4^{+16}_{-3} \, \%$ and an acceptance rate of $0.023 \, \%$. The assigned uncertainties are derived using parametric bootstrapping. We assume that measurements of a logical qubit in an arbitrary basis follow a binomial distributions characterized by probability $p_\mathrm{L}$ of measuring $1$ and number of shots $m$. New sets of measurement outcomes are drawn from a binomial distribution based on experimental estimates of $p_\mathrm{L}$ and $m$. This is repeated 1000 times and executed for each basis $X$, $Y$ and $Z$. The density matrix of the state and the corresponding fidelity is calculated from the bootstrapped data. The median and $1\sigma$-interval of the derived fidelity distribution are reported. The realization of a non-Clifford gate successfully concludes the universal gate set on the $[[5,1,3]]$ error correction code.

So far, all processes have been characterized in terms of logical state fidelity, which includes SPAM errors. In the next chapter, we move towards a process tomography removing SPAM noise and extracting the bare process fidelity.

\section{Characterization of logical qubit}\label{sec:characterization}

The performance of logical operations and QEC gadgets can be modeled as the composition of a quantum channel $\mathcal{E}_\text{op}$ that represents the ideal operation $\mathcal{E}_\text{ideal}$ with a completely positive trace-preserving (CPTP) map $\mathcal{E}_\text{noise}$ representing the noise on the actual gate being implemented
\begin{equation}
\mathcal{E}_\text{op}(\rho)=(\mathcal{E}_\text{noise}\circ\mathcal{E}_\text{ideal})(\rho).
\label{eq:noise-channel}
\end{equation}
The performance of a qubit is characterized by estimating $\mathcal{E}_\text{noise}$ using quantum process tomography (QPT)~\cite{chuang1997prescription}. Quantum channels $\mathcal{E}$ can be defined by a process matrix $\chi$
\begin{equation}
\mathcal{E}(\rho)=\sum_{i,j=0}^{4^N-1}\chi_{ij}P_i\rho P_j,
\end{equation}
where $N$ is the number of qubits and $P_i,P_j$ run over the $N$-qubit Pauli group $\mathcal{P}_N=\{I,X,Y,Z\}^{\otimes N}$. The process matrix is described by $4^N(4^N-1)$ independent real parameters. Preparing states in the set $\{\ket{0},\ket{1},\ket{+},\ket{+i}\}^{\otimes N}$, applying the operation and measuring them in all the combinations of Pauli bases estimates the parameters. This can be done by executing $4^N3^N$ independent circuits, what makes the experimental estimation of the full process matrix impractical for larger system sizes. To overcome this issue, we assume a Pauli-noise model, which has shown good agreement with experimental data in the context of QEC experiments~\cite{postler2022demonstration, postler2024demonstration, pogorelov2025experimental, gutierrez2015comparison, gutierrez2016errors}. The Pauli-noise model gives rise to a diagonal $\chi$ matrix describing $\mathcal{E}_\text{noise}$. Characterizing physical noise in a $[[5,1,3]]$ logical qubit requires the estimation of $4^5-1=1023$ parameters, which is still impractical for an experimental realization. 
We perform QPT at the logical level, instead of considering all physical degrees of freedom, to further reduce the number of parameters.

During syndrome extraction, physical Pauli errors are removed and converted to logical Pauli errors. If syndrome extraction is performed after every logical operation, the quantum channels can only affect the logical degrees of freedom and can thus be expressed as logical Pauli channels. This lowers the number of parameters to 4 for a single-qubit logical operation. Note however that syndrome extraction is not perfect and leaves uncorrected physical errors. These residual errors, that can combine with errors introduced during subsequent operations, increase the logical error probability with respect to the case of ideal syndrome extraction. Although logical process tomography does not provide a complete description for concatenated logical operations, it can be used to faithfully capture the performance of a single logical operation on an ideal logical qubit which has not yet suffered an error. However, one needs to separate the error of the operation from those of logical state preparation and readout.

Instead of the $\chi$ matrix, it is more convenient to work with Pauli transfer matrices (PTMs)~\cite{Chow2012Universal}
\begin{equation}
(R_\Phi)_{ij}=\textrm{tr}[P_i\mathcal{E}(P_j)].
\end{equation}
Under the PTM representation, the composition of two maps corresponds to the multiplication of PTMs, which eases the manipulation of quantum operations. For example, Eq.~\eqref{eq:noise-channel} can be written as
\begin{equation}
R_\text{op}=R_\text{noise}R_\text{ideal}.
\end{equation}
Since $\mathcal{E}_\text{op}$ (and thus $R_\text{op}$) can be estimated by QPT, the noise PTM $R_\text{noise}$ is extracted by inverting the ideal PTM.

\subsection{Logical SPAM error removal}
The experimentally estimated noise PTM $R_\text{est}$ is affected by SPAM errors and differs from the operation PTM $R_\text{op}$, which we want to characterize
\begin{equation}
R_\text{est}=R_\text{meas}R_\text{op}R_\text{prep}.
\end{equation}
It is common to estimate SPAM errors by performing QPT on encoding and readout of physical~\cite{Bantysh2019high,Samach2022Lindblad} or logical~\cite{Vezvaee2026surface} states, obtaining $R_\text{meas}R_\text{prep}$. In principle, state preparation errors are indistinguishable from measurement errors~\cite{Blume2025easy} due to the gauge freedom that also appears in the context of gate set tomography (GST)~\cite{nielsen2021gate,Blume2017demonstration}.
Usually, a noise model parametrized by a set of failure probabilities $\boldsymbol{p}$ and maximum likelihood estimator (MLE) is used to predict the channels $R_\text{meas}(\boldsymbol{p})$ and $R_\text{prep}(\boldsymbol{p})$ that get closest to the experimentally observed $R_\text{meas}R_\text{prep}$~\cite{Samach2022Lindblad}.

In the context of logical qubit tomography, we introduce in App.~\ref{app:qst-characterization} a model-independent scheme to characterize the full encoding circuit $R_\text{prep}$. However, the overhead of that protocol prevents its application for our current $[[5,1,3]]$ code experiments. Even if we cannot resolve the SPAM errors in preparation and measurements, we provide a principled estimation of the logical fidelity without SPAM contributions. Note that those processes act trivially on the encoded information in the noiseless quantum memory experiment, setting $R_\text{ideal}=I$. We obtain the entanglement fidelity~\cite{nielsen2002simple} of the channel without the SPAM contribution from the estimated matrix $R_\text{est}$, by noticing that
\begin{equation}
\begin{aligned}
F_e &= \frac{1}{4^N}\mathrm{tr}\left(R_\text{noise}\right) = \frac{1}{4^N}\mathrm{tr}\left[R_\text{meas}^{-1}R_\text{est}R_\text{prep}^{-1}\right]\\&=\frac{1}{4^N}\mathrm{tr}\left[(R_\text{meas}R_\text{prep})^{-1}R_\text{est}\right].
\end{aligned}
\label{eq:SPAM-removal-fidelity}
\end{equation}
From the entanglement fidelity, we obtain the average fidelity $F_\text{avg}=\frac{F_e+2^{-N}}{1+2^{-N}}$ of the noise channel~\cite{nielsen2002simple}.

As mentioned, the SPAM-error PTM $R_\text{meas}R_\text{prep}$ is determined experimentally by measuring the qubit immediately after the logical state preparation. The matrix is approximately inverted to obtain the fidelity of the process being characterized using Eq.~\eqref{eq:SPAM-removal-fidelity}. Due to shot noise, linear inversion of PTMs can give non-physical channels and typically requires maximum-likelihood approaches. Under a Pauli noise model, it is enough to ensure that all components of the PTM are positive, rendering MLE unnecessary.

Noise process matrices commute under the assumption of Pauli noise, which allows estimating the full matrix without the SPAM-error contribution
\begin{equation}
R_\text{noise}=R_\text{meas}^{-1}R_\text{est}R_\text{prep}^{-1}=(R_\text{meas}R_\text{prep})^{-1}R_\text{est}.
\label{eq:SPAM-removal-chi}
\end{equation}

The condition $R_\text{op}=I$, which we imposed in the ideal setting, does not hold for logical primitives that require postselection because the channel is no longer trace-preserving. In that situation, the SPAM-error removal protocol is not valid since we  require that $R_\text{noise}$ inserts noise instead of removing it. Thus, the protocol only works when the characterized operation includes error correction but not error detection. Since SPAM errors are really low when applying postselection, we can use the SPAM-error removal protocol when using error correction and keep the directly estimated PTM $R_\text{est}$ when working with postselection.

\begin{figure}
\centering
\includegraphics[width=.8\linewidth]{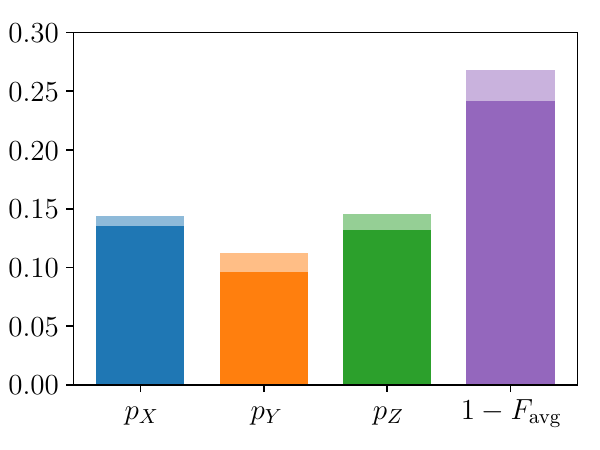}
\caption{\justifying \textbf{Logical process tomography of the quantum memory experiment.} We prepare a logical qubit, perform fault-tolerant syndrome extraction, apply corrections and fault-tolerantly measure. Using logical QPT, we obtain the logical process matrix of the quantum memory experiment assuming a Pauli noise model. $p_X$, $p_Y$ and $p_Z$ correspond to the diagonal entries of the logical process matrix and we additionally determine the average infidelity of the process. We calibrate SPAM errors by immediately measuring the logical qubit after encoding. The SPAM errors are removed (dark bars) from the raw tomography data (light bars).}
\label{fig:Syndrome-Extraction-Characterization}
\end{figure}

In Fig.~\ref{fig:Syndrome-Extraction-Characterization}, the results of the FT quantum memory experiment are shown including preparation, syndrome extraction, correction and measurement assuming logical Pauli noise and avoiding postselection beyond state preparation. The total logical noise channel (light bars) is compared to the SPAM free noise channel (dark bars). The effect of SPAM errors is eliminated using Eqs.~\eqref{eq:SPAM-removal-fidelity} and~\eqref{eq:SPAM-removal-chi}, increasing the average infidelity by $3\%$, which is comparable to the SPAM error characterized in Sec.~\ref{sec:toolbox} Fig.~\ref{fig:single_qubit_qec}(d).

So far, we have been considering a Pauli noise channel $R_\text{noise}$ appended after every logical operation. In the following subsection, we extend the protocol for non-Clifford gates, where the Pauli noise assumption is no longer valid.

\subsection{Non-Clifford gate characterization}
When characterizing non-Clifford gates, the assumption of Pauli noise breaks as Pauli errors are not mapped to Pauli operators after the gate. Instead, we model noise as two Pauli channels that get applied before and after the non-Clifford gate
\begin{align}
&R_\text{op}=R_\text{after}R_\text{ideal}R_\text{before}\\
&\implies R_\text{noise}=R_\text{after}(R_\text{ideal}R_\text{before}R_\text{ideal}^{-1}).
\end{align}
We apply this noise model to the characterization of the logical T gate. Looking at the structure of $R_\text{noise}$, we observe that the only non-trivial components are $R_{XX},R_{YY},R_{ZZ},R_{XZ},R_{ZX}$. We have to measure $\ket{0}_\mathrm{L}$ in the $X$ basis and $\ket{+}_\mathrm{L}$ in the $Z$ basis additionally compared to a logical Clifford characterization. The number of logical measurements required for this approach is still lower than the 9 measurements required for full process tomography. The resulting process matrix for the logical T gate is shown in Fig.~\ref{fig:results_magic_states}(d) achieving an $85 \pm 8 \, \%$ average gate fidelity.

\section{Discussion}\label{sec:dicsussion}

In this work, we have demonstrated a complete set of fault-tolerant logical primitives for a non-CSS code, the perfect $[[5, 1, 3]]$ code, on a trapped-ion quantum processor for the first time. 
The implemented toolbox comprises mid-circuit readout and real-time conditional control, allowing for logical state preparation, quantum error correction cycles on up to two logical qubits in parallel, logical measurements, logical entangling operations, magic-state preparation and injection, and modular logical state teleportation. Together, this suite of fault-tolerant operations supports all ingredients required for universal logical control with a non-CSS code. 

Similarly, large qLDPC codes will require to selectively and fault-tolerantly address individual logical qubits encoded in one block, which cannot be achieved with a global transversal scheme. 
Thus, the demonstration of these non-trivial fault-tolerant protocols becomes relevant not only for non-CSS codes, but also when working with scalable CSS qLDPC codes.

We have considered the smallest possible quantum error-correcting code capable of correcting an arbitrary single-qubit error. 
The combination of a non-fault-tolerant encoding, verification, and subsequent error detection has been explored theoretically for CSS codes~\cite{goto2016minimizing, chamberland2019fault, gidney2024magic, sahay2025fold} and used in experimental demonstrations of fault-tolerant logical state preparation and the implementation of logical gates~\cite{lacroix2025scaling, butt2026demonstration, pogorelov2025experimental, postler2022demonstration, rosenfeld2025magic}. 
Related ideas appear in magic-state cultivation~\cite{sahay2025fold, gidney2024magic}, where a verified magic state is prepared on a small code and subsequently grown into a larger code through interleaved unitary transformations and stabilizer measurements. The perfect five-qubit code offers a more compact alternative to conventionally used color codes and rotated surface codes and may reduce the required qubit count for these types of protocols. 
Our results therefore offer a near-term alternative to larger distance-three CSS codes, while retaining the ability to perform and optimize universal logical operations. 
This compactness makes the code a useful testbed for exploring fault-tolerant protocols beyond the CSS setting and for assessing how such methods can be realized experimentally.

A central element of the demonstrated toolbox is modular logical state teleportation~\cite{Nielsen_and_Chuang, horsman2012surface}. 
This primitive transfers quantum information between encoded blocks through logical measurements and real-time feedback, without physically moving or directly coupling logical qubit modules. 
It therefore provides a demonstration of a key requirement for many modular scalable fault-tolerant architectures, such as interconnected refrigerators for superconducting qubits \cite{ibm_blog, ibm_hw} and trapped-ion QCCD architectures \cite{quantinuum_hw, ionq_hw}.
Our implementation shows that such connectivity primitives can be realized in a compact non-CSS code, extending techniques that are more naturally supported in CSS-code architectures.

Our protocols, here implemented for the five-qubit code, show that non-CSS codes can support the full set of FT primitives, including teleportation-based connectivity and non-Clifford resource injection, which are needed for experimentally viable modular fault-tolerant quantum computing. This expands the experimental toolkit for fault-tolerant quantum computation beyond the CSS setting and provides a route toward compact modular architectures based on more general quantum error-correcting codes.

\section*{Acknowledgements}
We gratefully acknowledge support by the European Union’s Horizon Europe research and innovation program under Grant Agreement Number 101114305 (“MILLENION-SGA1” EU Project) (A.B, C.B, T.M., M.Meyer, M.M\"{u}ller), the US Army Research Office through Grant Number W911NF-21-1-0007 (F.B., T.M., M.M\"{u}ller), the Austrian Research Promotion Agency under Contract Number 897481 (HPQC) (T.M.) supported by the European Union – NextGenerationEU, the Austrian Science Fund (FWF Grant-DOI 10.55776/F71) (SFB BeyondC) (T.M.), the  Spanish Ministry of Science, Innovation and Universities (FPU24/01105) (C.B.), QUITEMAD-CM TEC-2024/COM-84 (A.B., C.B.), the Grant IFT Centro de Excelencia Severo Ochoa CEX2020-001007-S (A.B., C.B.) funded by MCIN/AEI/10.13039/501100011033, and Intelligence Advanced Research Projects Activity (IARPA), under the Entangled Logical Qubits program through Cooperative Agreement Number W911NF-23-2-0216. 
We further receive support from the IQI GmbH, and by the Deutsche Forschungsgemeinschaft (DFG, German Research Foundation) under Germany’s Excellence Strategy Cluster of Excellence Matter and Light for Quantum Computing (ML4Q) EXC 2004/1 390534769 and through the DFG Priority Programme SPP
2514. This research is also part of the Munich Quantum Valley (K-8), which is supported by the Bavarian state government with funds from the Hightech Agenda Bayern Plus.

The views and conclusions contained in this document are those of the authors and should not be interpreted as representing the official policies, either expressed or implied, of IARPA, the Army Research Office, or the U.S. Government. The U.S. Government is authorized to reproduce and distribute reprints for Government purposes notwithstanding any copyright notation herein.

We acknowledge computing time provided at the NHR Center NHR4CES at RWTH Aachen University (Project No. p0020074) (F.B., M.M\"{u}ller). This is funded by the Federal Ministry of Education and Research and the state governments participating on the basis of the resolutions of the GWK for national high-performance computing at universities.

\section*{Data availability}
The data provided in the figures in this article, the explicit circuits and the code that was used to simulate the presented protocols are available at \url{https://doi.org/10.5281/zenodo.22796457}.  

\section*{Author contributions:}
R. F. implemented the presented protocols on the experimental setup.
F. B. developed the presented protocols and performed simulations. C. B. developed and applied the logical process tomography. R. F., I. P., M. Me. and A. S. built and maintained the experimental setup. R. F., F. B. and C. B. analyzed results. R. F., F. B., C. B., I. P., and M. Me. wrote the manuscript with contributions from all authors. A. B., M. Mü. and T. M. supervised the project.

\clearpage
\appendix
\section{Numerical methods}\label{app:numerical_methods}

We simulate noisy quantum circuits by performing Monte Carlo simulations. 
Each noisy circuit component is modeled by first applying the respective ideal operation, followed by an error $E$ occurring with probability $p$. 
We simulate a depolarizing noise channel after every single- and two-qubit gate
\begin{align}
    \mathcal{E}_1(\rho) &= (1 - p_1)\rho + \frac{p_1}{3} \sum_{j= 1}^3 E^{j}_1 \rho E^{j}_1, \label{eq:depol_single_qubit} \\
    \mathcal{E}_2(\rho) &= (1 - p_2)\rho + \frac{p_2}{15} \sum_{j= 1}^{15}  E_2^{j} \, \rho\, E_2^{j}. \nonumber
\end{align}
Here, an error is applied with probability $p_1$ and $p_2$ from the error sets $E_1$ and $E_2$ defined by
\begin{align}
	E_1 &\in \{ \sigma_k, \forall k \in \{1, 2, 3 \} \}, \\
	E_2 &\in \{\sigma_k \otimes \sigma_l, \forall k, l \in \{0, 1, 2, 3 \}  \} \backslash  \{\sigma_0 \otimes \sigma_0 \}, \nonumber
\end{align}
where $\sigma_k$ are the single-qubit Pauli operators \hbox{$\sigma_k = \{I, X, Y, Z \}$ with $k=0, 1, 2, 3$}. 
All qubits are initialized and measured in the $Z$ basis. We simulate faults in these operations by applying $X$-errors after initialization and before measurement, each occurring with probabilities $p_{\mathrm{init}}$ and $p_{\mathrm{meas}}$, respectively. 

Moreover, we include dephasing on idle qubits modeled with the error channel~\cite{heussen2023strategies}
\begin{align}
    \mathcal{E}_{\mathrm{idle}}(\rho) &= (1 - p_{\mathrm{idle}})\rho + p_{\mathrm{idle}} Z\rho Z. 
\end{align}
The probability $p_{\mathrm{idle}}$ indicates the likelihood of a $Z$-error on an idling qubit during the execution of a gate on a different qubit. This error probability depends on the execution time $t$ of the performed gate and the qubit coherence time $T_2$~\cite{pal2022relaxation}
\begin{align}
    p_{\mathrm{idle}} = \frac{1}{2} \left[1 - \mathrm{exp}\left(-\frac{t}{T_2} \right) \right]. 
\end{align}

The simulations use the physical error rates determined in the experiment as summarized in Tab.~\ref{tab:error_rates_simulation}. 
Lastly, noise on each idling qubit during a mid-circuit measurement of another qubit is modeled as an asymmetric single-qubit depolarizing channel
\begin{align}
    \mathcal{E}_{\mathrm{mcm}}(\rho) &= \left[1 - (p_x + p_y + p_z) \right]\rho \\ &+ p_x X \rho X + p_y Y \rho Y + p_z Z \rho Z\nonumber
\end{align}
with distinct error rates $p_x, p_y, p_z$~\cite{postler2024demonstration}. In simulations, the three values are set to the diagonal elements of the process matrix characterized in Fig.~\ref{fig:midcirucit_processmatrix}.

\begin{table}[!ht]
    \centering
    \renewcommand*{\arraystretch}{1.2}
    \caption{\justifying \textbf{Error rates and durations of operations on the trapped-ion quantum processor~\cite{butt2026demonstration, pogorelov2025experimental, postler2024demonstration}. }We use these probabilities in all presented numerical simulations of quantum circuits on the trapped-ion quantum processor. }
    \begin{tabular}{|c|c|c|}
        \hline
         Operation & Error rate & Duration\\
         \hline
         Two-qubit gate & $p_{2} = 0.02$ & \SI{350}{\micro\second}\\
         \hline
         Single-qubit gate & $p_{1} = 0.001$ & \SI{70}{\micro\second}\\
         \hline
         Measurement & $p_{\mathrm{meas}} = 0.003$& \SI{1}{\milli\second}\\
         \hline
         Preparation & $p_{\mathrm{init}} = 0.003$& \SI{100}{\micro\second}\\
         \hline
		 \hline
         Coherence time &  \multicolumn{2}{c|}{$T_2 = \SI{200}{\milli\second}$}  \\
        \hline
    \end{tabular}
    \label{tab:error_rates_simulation}
\end{table}

\section{Mid-circuit measurements} \label{app:Mid-circuit_process}

\begin{figure*}
    \centering
    \includegraphics[width=180mm]{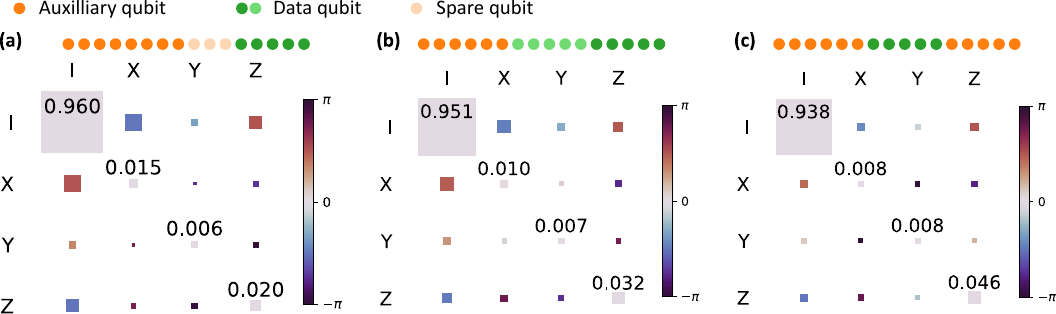}
    \caption{\justifying \textbf{Process matrix of a mid-circuit measurement.} The qubit state of the data qubits is maintained during a mid-circuit measurement while the auxiliary qubits are detected and reset. Due to infidelities, the data qubits experience a non-identity process, which is characterized with single-qubit process tomography and the results are averaged across all data qubits. (a) One logical qubit is encoded in the qubit register using 8 auxiliary qubits and the corresponding process tomography is shown. We use two other configurations: (b) encoding two logical qubit with the support of 6 auxiliary qubits and (c) encoding a single logical qubit assisted by 11 auxiliary qubits.}
    \label{fig:midcirucit_processmatrix}
\end{figure*}

The qubit register is split into two parts during mid-circuit measurements. The data qubits keep their quantum state during a mid-circuit measurement, while the auxiliary qubits are measured and reset to $\ket{0}$. In an ideal case, the process tomography of a mid-circuit measurement across the data qubits would yield an identity process. Stray light from the measurement, dephasing during idling periods and pulse infidelities introduce errors on the data qubits. 

Three different partitions of the ion string are used. If only a single logical qubit is encoded in the qubit register, the first 8 ions are used as auxiliary qubits and the last 5 ions are used as data qubits. The three inner ions are not utilized in this scenario and serve as a buffer. If two logical qubits are encoded in the ion chain, the first six ions are auxiliary qubits and the last 10 ions are data qubits. The fault-tolerant magic state injection makes it necessary to measure one logical qubit out of two. This measurement is realized by assigning the first 6 ions as auxiliary qubits, the next 5 ions as data qubits and the last 5 ions as auxiliary qubits as well. Figure~\ref{fig:midcirucit_processmatrix} illustrates the partition of the ion string at the top. Single-qubit process tomography of the mid-circuit measurement is performed on the data qubits and the average is shown below the respective partitioning in Fig.~\ref{fig:midcirucit_processmatrix}.

The average process fidelities shown in 
Fig.~\ref{fig:midcirucit_processmatrix} are higher compared to the results in Ref.~\citenum{postler2024demonstration}. This has been achieved by increasing the coherence time of the optical encoding ($\ket{0}=\ket{4 ^2\textrm{S}_{1/2}, m_J = -1/2}$ to $\ket{1}=\ket{3 ^2\textrm{D}_{5/2}, m_J = -1/2}$) from roughly \SI{50}{\milli\second} to \SI{200}{\milli\second}, the ground state encoding ($\ket{0}=\ket{4 ^2\textrm{S}_{1/2}, m_J = -1/2}$ to $\ket{1}=\ket{4 ^2\textrm{S}_{1/2}, m_J = +1/2}$) from approx. \SI{5}{\milli\second} to \SI{100}{\milli\second} and the encoding within the D level ($\ket{0}=\ket{3 ^2\textrm{D}_{5/2}, m_J = -1/2}$ to $\ket{1}=\ket{3 ^2\textrm{D}_{5/2}, m_J = +1/2}$) from around \SI{8}{\milli\second} to \SI{100}{\milli\second} by improving magnetic field stability.

\section{Flag QEC on the $[[5, 1, 3]]$ code}
\label{app:FlagQEC}
\begin{figure*}
    \centering
    \includegraphics[width=170mm]{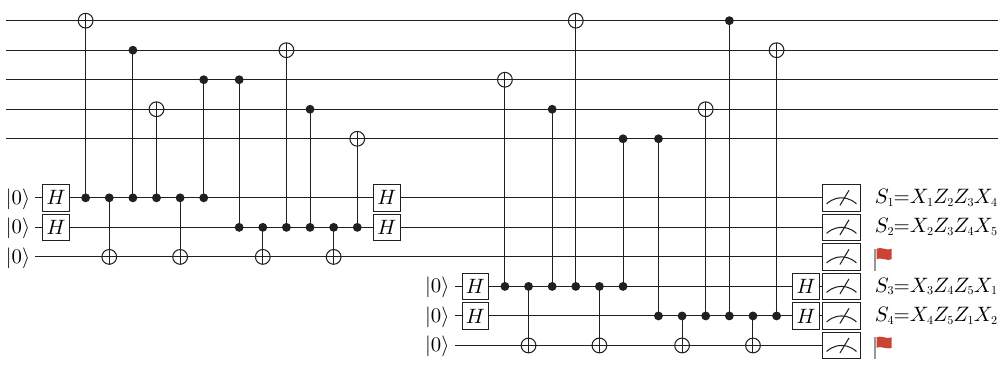}
    \caption{\justifying \textbf{Fault-tolerant stabilizer measurement.} We first measure the two stabilizers $S_1$ and $S_2$ with a single flag in parallel~\cite{reichardt2020fault}, followed by a second parallelized stabilizer measurement of $S_3$ and $S_4$. }
    \label{fig:stabilizer_measurement}
\end{figure*}

We use the circuit shown in Fig.~\ref{fig:stabilizer_measurement} to measure all four stabilizers fault-tolerantly. We implement the following protocol for one round of FT QEC on the $[[5, 1, 3]]$ code:
\begin{enumerate}
    \item We measure all stabilizers with the circuit shown in Fig.~\ref{fig:stabilizer_measurement}. 
    \item \begin{enumerate}
        \item If all flags have been in $|0\rangle$ and the syndrome is (0, 0, 0, 0): do nothing. 
        \item If all flags have been in $|0\rangle$ and the syndrome is non-trivial: Measure all stabilizers again without flags. Run the standard LUT~\ref{tab:LUT_QEC_standard}. 
        \item Else, a flag has been in $|1\rangle$: Measure all stabilizers again without flags. If the new syndrome matches one in the flag-LUT~\ref{tab:LUT_QEC_parallel}, apply the respective flag-error. If it does not match any in this table, run the standard LUT~\ref{tab:LUT_QEC_standard}. 
    \end{enumerate}
\end{enumerate}

\section{Fault-tolerant logical state teleportation}\label{app:detailed_FT_teleportation_protocol}

The full protocol for FT logical state teleportation is as follows. 

\begin{figure*}
    \centering
    \includegraphics[width=126mm]{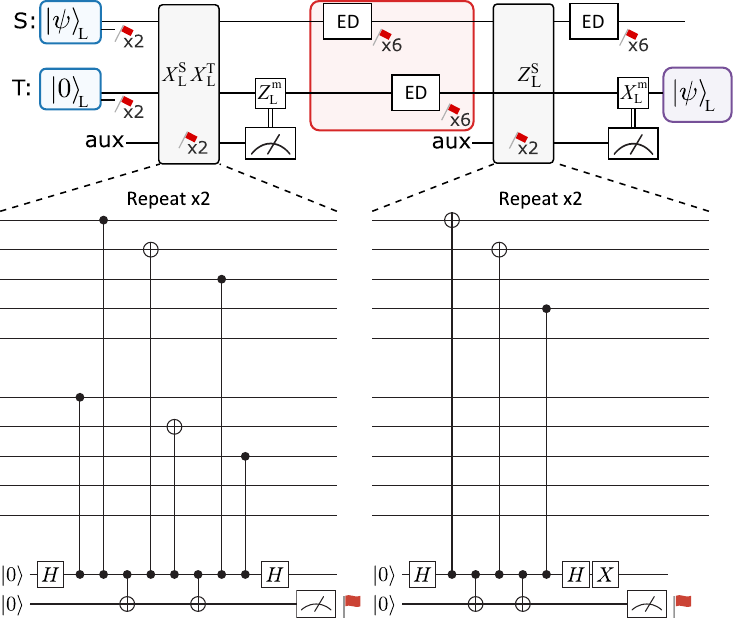}
    \caption{\justifying \textbf{Circuit for fault-tolerant logical state teleportation.} After preparing the target register in $|0\rangle_\mathrm{L}$, we measure the joint logical operator $X_\mathrm{L}^\mathrm{S} X_\mathrm{L}^\mathrm{T}$ twice with a single flag each. The measurement outcomes determine the application of a logical Pauli feedback operation to the target register. After a round of error detection on each logical qubit register, we similarly measure $Z_\mathrm{L}^\mathrm{S}$ two times with flag qubits. We finally apply $X_\mathrm{L}^\mathrm{m}$ to the target register, depending on the measurement outcome $m$. }
    \label{fig:teleportation_detailed}
\end{figure*}

\begin{enumerate}
    \item \textbf{Fault-tolerant measurement of $X_\mathrm{L}^\mathrm{S} X_\mathrm{L}^\mathrm{T}$.}
    We measure
    \begin{equation}
        X_\mathrm{L}^\mathrm{S} X_\mathrm{L}^\mathrm{T}
        =
        Z^\mathrm{S}_1 X^\mathrm{S}_2 Z^\mathrm{S}_3
        Z^\mathrm{T}_1 X^\mathrm{T}_2 Z^\mathrm{T}_3
    \end{equation}
    twice using the flagged circuit in the first half of Fig.~\ref{fig:teleportation_detailed}. Let the two outcomes be $m_1,m_2\in\{\pm1\}$.
    \begin{itemize}
        \item If a flag is raised, we perform one round of unflagged QEC on both logical blocks, correct the flagged error according to the lookup table in Tab.~\ref{tab:Flag_LUT_joint_X_measurement}, and then remeasure $X_\mathrm{L}^\mathrm{S} X_\mathrm{L}^\mathrm{T}$ once without flags.
        \item If no flag is raised and $m_1=m_2$, we perform one round of flagged QEC on both the source and target blocks. If an error is detected, we correct it and remeasure $X_\mathrm{L}^\mathrm{S} X_\mathrm{L}^\mathrm{T}$ once without flags. If no error is detected, we accept the repeated value $m_1=m_2$.
        \item If no flag is raised and $m_1\neq m_2$, we perform one round of unflagged QEC on both logical blocks, correct any detected errors, and remeasure $X_\mathrm{L}^\mathrm{S} X_\mathrm{L}^\mathrm{T}$ once without flags.
    \end{itemize}
    If the final accepted outcome is $-1$, we apply the correction $Z_\mathrm{L}^\mathrm{T}$.

    \item \textbf{Fault-tolerant measurement of $Z_\mathrm{L}^\mathrm{S}$.}
    We then measure the source logical operator
    \begin{equation}
        Z_\mathrm{L}^\mathrm{S} = -X^\mathrm{S}_1 X^\mathrm{S}_2 Z^\mathrm{S}_4
    \end{equation}
    twice using the flagged circuit in the second half of Fig.~\ref{fig:teleportation_detailed}. Let the two outcomes be $n_1,n_2\in\{\pm1\}$.
    \begin{itemize}
        \item If a flag is raised, we perform one round of unflagged QEC on the source block, correct the flagged error according to the lookup table in Tab.~\ref{tab:Flag_LUT_Z_measurement}, and then remeasure $Z_\mathrm{L}^\mathrm{S}$ once without flags.
        \item If no flag is raised and $n_1=n_2$, we perform one round of flagged QEC on the source block. If an error is detected, we correct it and remeasure $Z_\mathrm{L}^\mathrm{S}$ once without flags. If no error is detected, we accept the repeated value $n_1=n_2$.
        \item If no flag is raised and $n_1\neq n_2$, we perform one round of unflagged QEC on the source block, correct any detected errors, and remeasure $Z_\mathrm{L}^\mathrm{S}$ once without flags.
    \end{itemize}
    If the final accepted outcome is $-1$, we apply the correction $X_\mathrm{L}^\mathrm{T}$.
\end{enumerate}

\begin{figure*}
    \centering
    \includegraphics[width=170mm]{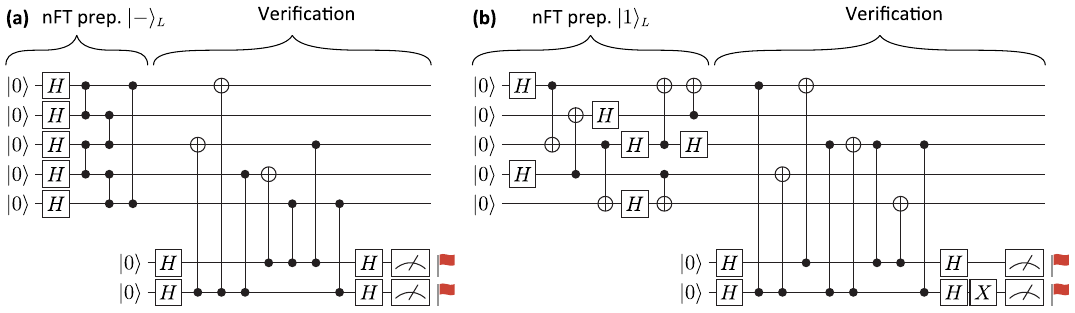}
    \caption{\justifying \textbf{Circuits for fault-tolerant logical state preparation. }We prepare (a) $|-\rangle_\mathrm{L}$ and (b)$|1\rangle_\mathrm{L}$ by performing a non-FT encoding, followed by a verification~\cite{zen2024quantum, ryan2022implementing}. The verification step consists of the measurement of two complementary representations of logical operators.}
    \label{fig:ft_init}
\end{figure*}

\begin{figure*}
    \centering
    \includegraphics[width=170mm]{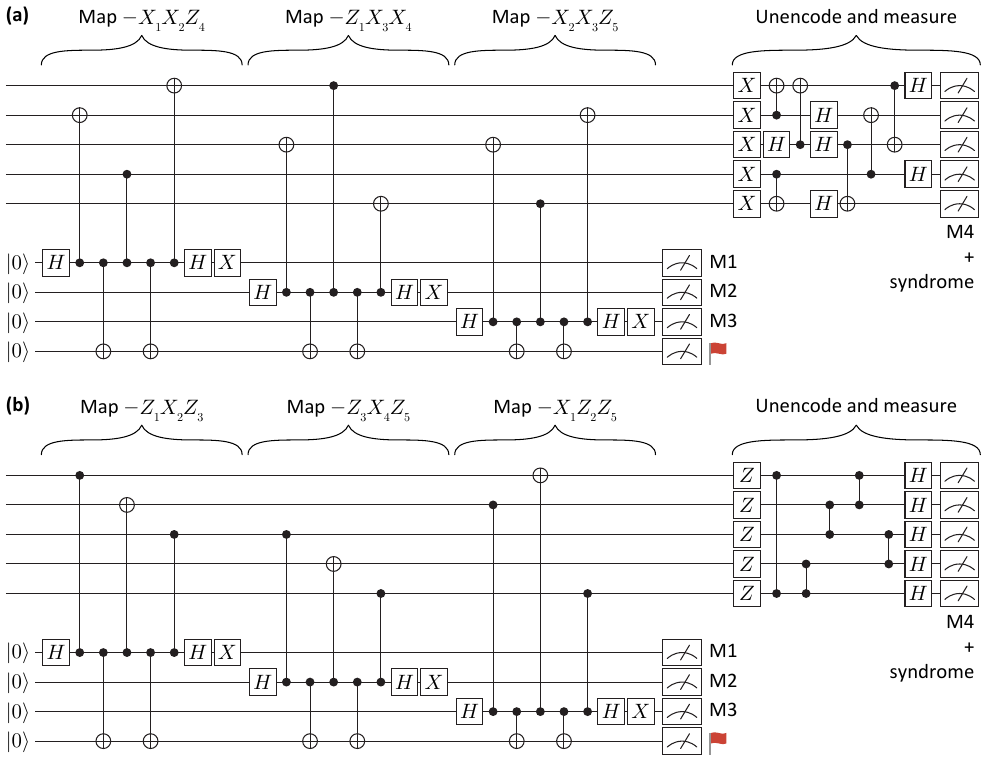}
    \caption{\justifying \textbf{Fault-tolerant logical measurements of (a) $Z_\mathrm{L}$ and (b) $X_\mathrm{L}$.} Each logical measurement consists of mapping three different representations of the respective logical operator onto a physical qubit that is coupled to an additional flag qubit. At the end, the logical state is unencoded using the inverted state preparation circuit and measured in the $Z$ basis. From this measurement, we can infer the logical value and a syndrome. }
    \label{fig:circuit_measure_logicals}
\end{figure*}

\begin{figure*}
    \centering
    \includegraphics[width=170mm]{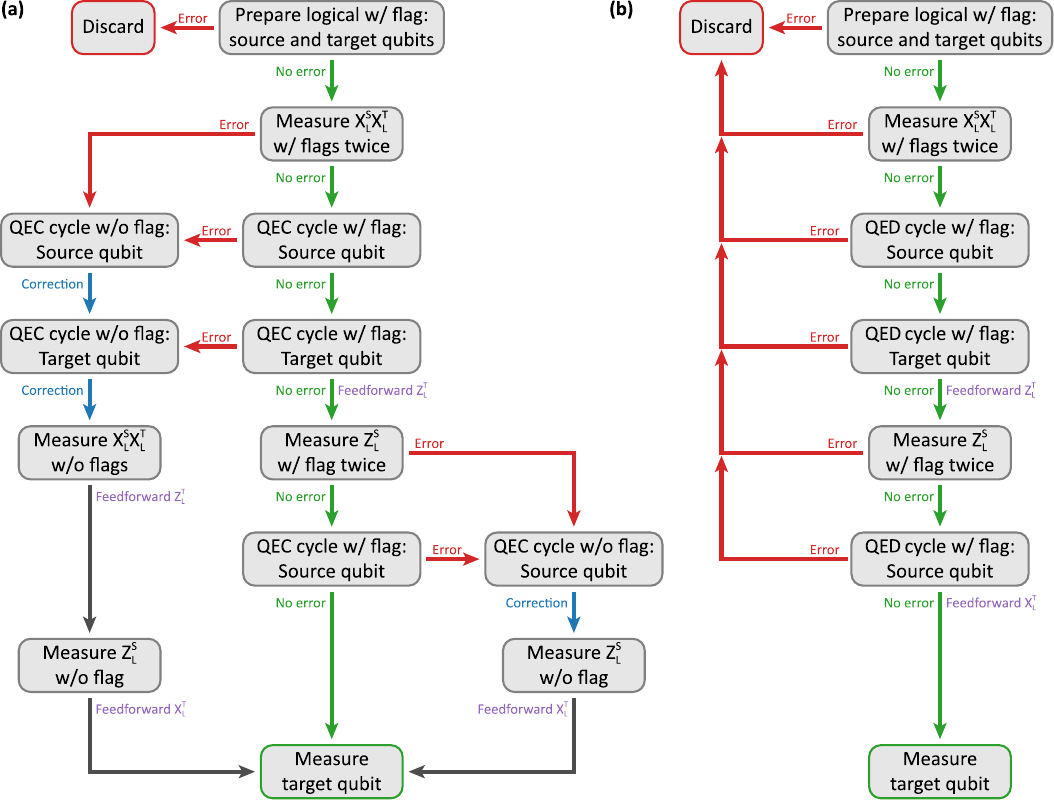}
    \caption{\justifying \textbf{Decision tree for fault-tolerant logical state teleportation. }(a) For fully error-correcting logical state teleportation, we start by preparing the logical source and target qubits. Then, the joint logical operator $X_\mathrm{L}^\mathrm{S} X_\mathrm{L}^\mathrm{T}$ is measured repeatedly with flags. Depending on the detection of an error, we perform flagged or unflagged QEC cycles on the source and target registers, followed by the measurement of $Z_\mathrm{L}^\mathrm{S}$ and a logical measurement of the target qubit. (b) For non-deterministic error-detecting logical state teleportation, we discard a run whenever any error is detected.}
    \label{fig:protocol}
\end{figure*}

Figure~\ref{fig:protocol}(a) shows the decision tree of the full teleportation in the deterministic error-correction case, which can be compressed substantially in the error-detection case to the scheme shown in Fig.~\ref{fig:protocol}(b).

\section{Fault-tolerant magic-state injection}\label{app:MS_injection_details}
\begin{figure*}
    \centering
    \includegraphics[width=170mm]{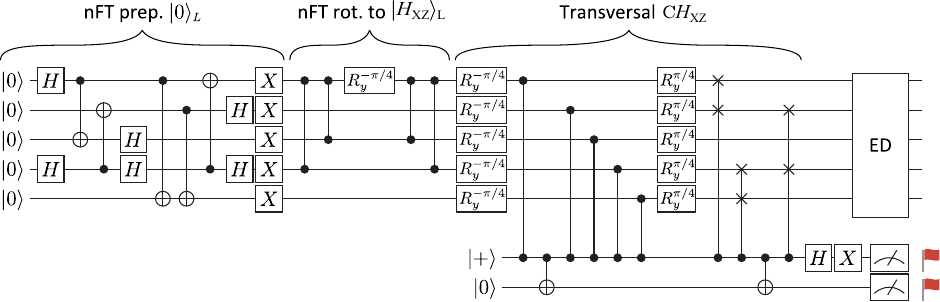}
    \caption{\justifying \textbf{Fault-tolerant magic-state preparation on the $[[5, 1, 3]]$ code. }First, the magic state \hbox{$|H_{XZ}\rangle_\mathrm{L} = \cos{({\frac{\pi}{8}})} |0\rangle_\mathrm{L} + \sin{({\frac{\pi}{8}})} |1\rangle_\mathrm{L}$} is prepared non-fault-tolerantly. We subsequently measure the logical Hadamard operator with a flag qubit and perform one round of error detection by measuring all four stabilizers once.}
    \label{fig:nft_h}
\end{figure*}

Figure~\ref{fig:nft_h} shows the full circuit for the FT preparation of a magic state $|H_{XZ}\rangle_\mathrm{L}$ on the $[[5, 1, 3]]$ code. 
The fault-tolerant measurement of $H_\mathrm{L}$ requires flag qubits. Although the logical Hadamard is implemented by transversal Hadamards together with a fixed qubit permutation, a controlled-$H_\mathrm{L}$ measurement also involves controlled-SWAP operations. Since these controlled-SWAPs share targets, a single fault can propagate to a weight-three data error, which is already a logical operator for the distance-three code. We therefore use a flagged measurement circuit to detect such dangerous error propagation. 

\begin{figure}
    \centering
    \includegraphics[width=85mm]{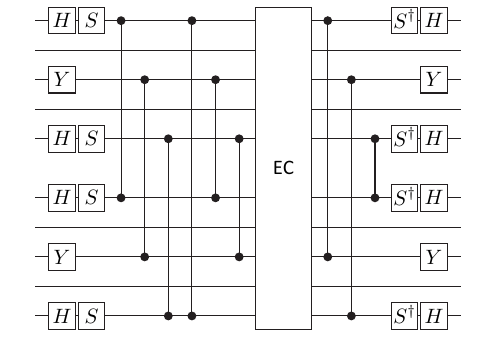}
    \caption{\justifying \textbf{Fault-tolerant logical entangling gate. } The C$Z_\mathrm{L}$ gate is implemented by means of a round-robin construction and an intermediate round of stabilizer extraction~\cite{ryan2022implementing, yoder2016universal}. }
    \label{fig:ft_cz}
\end{figure}

Figure~\ref{fig:ft_cz} shows the full circuit of the round-robin logical C$Z_\mathrm{L}$ gate between two $[[5, 1, 3]]$ codes. In particular, each stabilizer generator contains an even number of $X$ or $Y$ operators on the qubits participating in the physical $CZ$ gates, namely qubits $1$, $3$, and $5$ in each block. Since $X$ and $Y$ components propagate through $CZ$ gates, these contributions occur in pairs and cancel in the stabilizer propagation.

After two rounds of the round-robin circuit, the stabilizer group is generated by
\begin{align}
    \big\langle
        - Z_1 Z_2 Z_3 X_4, 
        &- X_2 Z_3 Z_4 Z_5, \nonumber \\
        - Z_6 Z_7 Z_8 X_9, 
        &- X_7 Z_8 Z_9 Z_{10}, \nonumber\\
        - Z_1 X_3 Z_4 Y_5 Z_6 Z_{10}, 
        &- Y_1 X_2 X_4 Y_5 Z_6 Z_8, \nonumber \\
        - Z_1 Z_5 Z_6 X_8 Z_9 Y_{10}, 
        &- Z_1 Z_3 Y_6 X_7 X_9 Y_{10}
    \big\rangle .
    \label{eq:generators}
\end{align}
Here qubits $1,\ldots,5$ belong to the first code block and qubits $6,\ldots,10$ to the second. The first four generators in Eq.~\eqref{eq:generators} remain confined to their original code blocks and are not affected by the preceding $CZ$ gates. We refer to them as the \emph{constant} stabilizer generators. 

\section{Error budget}
\label{app:ErrorBudget}

The numerical simulations make use of an experimentally informed noise model to generate data for comparison. The parameters of the noise model are described in App.~\ref{app:numerical_methods}. Assuming that the noise model of the simulation  approximately captures the experimental imperfections, we deduce an error budget by separating different error sources. The error sources are resolved by keeping a single error rate and setting all error probabilities to 0. The infidelity from two-qubit gates, coherence time and mid-circuit measurements are most dominant and we do not resolve the error contributions from single-qubit gates, measurement and preparation. 
The infidelity is averaged over the three Pauli states, if necessary. The infidelity of the three error sources do not sum up to the total infidelity because the errors influence each other, increasing the final error rate. 

\begin{figure*}
    \centering
    \includegraphics[width=180mm]{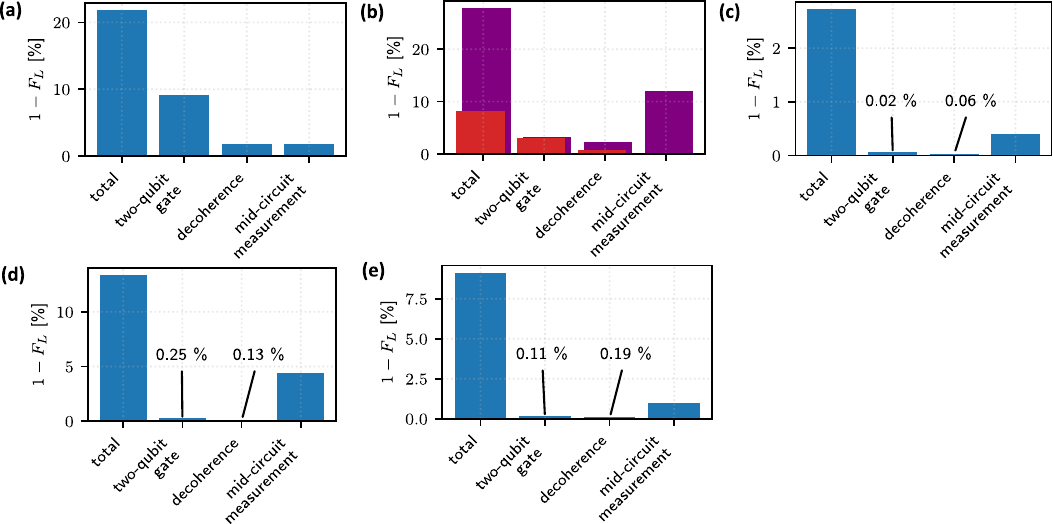}
    \caption{\justifying \textbf{Error budget. }Experimentally informed simulations enable the separation of error sources. Repeating simulations with only a single non-zero error rate gives an estimate of the infidelity contribution from this error source. (a) Error budget of the quantum memory experiment presented in Sec.~\ref{sec:toolbox} Fig.~\ref{fig:single_qubit_qec}(e). The infidelity is mainly limited by two-qubit gate errors and errors from decoherence and mid-circuit measurements contribute roughly equally. (b) Error contributions for the logical spectator error experiment presented in Sec.~\ref{sec:toolbox} Fig.~\ref{fig:results_crosstal}(b) are shown. The red bars in front correspond to the infidelity preparing two logical qubits and measuring one of them. The purple bars in the background represent the error budget of the spectator qubit after performing a QEC cycle. (c) The error budget of the preparation of two logical qubits and performing QED on both is shown corresponding to Sec~\ref{sec:toolbox} Fig.~\ref{fig:results_teleportation}(c). (d) The error budget of the state teleportation presented in Sec.~\ref{sec:toolbox} Fig.~\ref{fig:results_teleportation}(d) is shown. (e) The error contributions while applying a C$Y_\mathrm{L}$ gate to the $\ket{+, 0}_{\mathrm{L}}$ input state corresponding to the results presented in Sec.~\ref{sec:toolbox} Fig.~\ref{fig:results_magic_states}(c) are shown.}
    \label{fig:error_budget}
\end{figure*}

Figure~\ref{fig:error_budget}(a) shows the error budget of the quantum memory experiment (Sec.~\ref{sec:toolbox}, Fig.~\ref{fig:single_qubit_qec}(e)). The dominant error contribution is two-qubit gates followed by decoherence errors and mid-circuit measurement errors, that both contribute equally to the overall infidelity.

The error budget for the logical spectator error is shown in Fig.~\ref{fig:error_budget}(b). The bars in the front correspond to the preparation of two logical qubits and measuring a single one, equivalent to the bars in the front in Sec.~\ref{sec:toolbox} Fig.~\ref{fig:results_crosstal}(b). The error contribution considering only noisy mid-circuit measurements is zero, since there are no mid-circuit measurements in the protocol. The bars in the back correspond to the infidelity of the idling qubit after performing a QEC cycle on the other logical qubit, equivalent to the bars in the background of Sec.~\ref{sec:toolbox} Fig.~\ref{fig:results_crosstal}(b). The increase in infidelity is mainly driven by mid-circuit measurement errors and small contributions of decoherence errors.

The error budget presented in Fig.~\ref{fig:error_budget}(c) corresponds to the infidelities presented in Sec.~\ref{sec:toolbox} Fig.~\ref{fig:results_teleportation}(c), where two logical qubits are prepared and error detection is run on both. The two previous error profiles characterize sequences, which both use error correction. Switching to error detection changes the contributions of different types of errors slightly. The two-qubit gate error contribution is lower and the main error source are mid-circuit measurements. The error contribution from decoherence is nearly negligible. The mid-circuit measurement errors mainly stem from imperfect transfer between two different electronic state encodings to protect the data qubits from the readout and subsequent cooling light and may be reduced by adding single-ion addressing laser beams \cite{chen2026noninvasivemcm} or by moving to a different trap architecture, so detection takes place in a different trapping potential that is displaced from the data qubits \cite{pino2021qccd}.

The infidelity for logical state teleportation is shown in Sec.~\ref{sec:toolbox} Fig.~\ref{fig:results_teleportation} and the corresponding error profile is similar to the latter, which is shown in Fig.~\ref{fig:error_budget}(d).

The error contributions for the C$Y_\mathrm{L}$ gate applied to the input state $\ket{+, 0}_{\mathrm{L}}$ are shown in Fig.~\ref{fig:error_budget}(e) and yield a similar profile as the previous two. Resolving the error budget of the magic state preparation has a substantially larger time overhead than the other simulations due to the usage of non-Clifford gates. Since the error budgets for the latter three experiments using error detection look similar, we expect a comparable structure for the preparation of a magic state.

\section{Lookup tables}
\label{app:LookupTables}

Tables~\ref{tab:LUT_QEC_standard}, \ref{tab:LUT_QEC_parallel}, \ref{tab:Flag_LUT_joint_X_measurement} and \ref{tab:Flag_LUT_Z_measurement} summarize the corrections to be applied to physical qubits based on the stabilizer and logical operator measurement outcomes in the presented FT logical state teleportation protocols. 

\begin{table}[!tb]
    \centering
    \renewcommand*{\arraystretch}{1.7}
    \caption{\justifying \textbf{Lookup table for the $[[5, 1, 3]]$ code. }Each single-qubit error (left) is assigned a syndrome (right). Whenever a syndrome is measured, we apply the respective single-qubit correction. }
    \begin{tabular}{|c|c|c|}
    \hline
     Error & Syndrome $(S_1, S_2, S_3, S_4)$\\
     \hline
     $X_1$ & (0, 0, 0, 1) \\
     $X_2$ & (1, 0, 0, 0) \\
     $X_3$ & (1, 1, 0, 0) \\
     $X_4$ & (0, 1, 1, 0) \\
     $X_5$ & (0, 0, 1, 1) \\
     \hline
     $Z_1$ & (1, 0, 1, 0) \\
     $Z_2$ & (0, 1, 0, 1) \\
     $Z_3$ & (0, 0, 1, 0) \\
     $Z_4$ & (1, 0, 0, 1) \\
     $Z_5$ & (0, 1, 0, 0) \\
     \hline
     $Y_1$ & (1, 0, 1, 1) \\
     $Y_2$ & (1, 1, 0, 1) \\
     $Y_3$ & (1, 1, 1, 0) \\
     $Y_4$ & (1, 1, 1, 1) \\
     $Y_5$ & (0, 1, 1, 1) \\
     \hline
    \end{tabular}
    \label{tab:LUT_QEC_standard}
\end{table}

\begin{table*}[!tb]
    \centering
    \renewcommand*{\arraystretch}{1.7}
    \caption{\justifying \textbf{Flag QEC lookup table for parallel stabilizer readout using circuit shown in Fig.~\ref{fig:stabilizer_measurement}. } }
    \begin{tabular}{|c|c|c|}
    \hline
     Raised flag qubit & flag error & resulting syndrome afterwards $(S_1, S_2, S_3, S_4)$\\
     \hline
     1 & $Z_3 X_4$ & (0, 1, 0, 0) \\
     & $Y_2 Z_3 X_4$ & (1, 0, 0, 1) \\
     & $X_2 Z_3 X_4$ & (1, 1, 0, 0) \\
     & $Z_3 Z_4$ & (1, 0, 1, 1) \\
     & $Z_3 Y_4$ & (1, 1, 0, 1) \\

     & $Z_2 Z_4 X_5$ & (1, 1, 1, 1)\\
     & $Y_2 Z_4 X_5$&  (0, 1, 1, 1)\\
     & $Z_4 X_5$ & (1, 0, 1, 0) \\
     & $X_4 X_5$ & (0, 1, 0, 1) \\
     & $Y_4 X_5$ & (1, 1, 0, 0) \\
     \hline
     2& $Z_5 X_1$ & (0, 1, 0, 1) \\
     & $Y_4 Z_5 X_1$ & (0, 0, 1, 1) \\
     & $X_4 Z_5 X_1$ & (1, 0, 1, 0) \\
     & $Z_5 Z_1$ & (1, 1, 1, 0) \\
     & $Z_5 Y_1$ & (1, 1, 1, 1) \\

     & $Z_4 Z_1 X_2$ & (1, 0, 1, 1)\\
     & $Y_4 Z_1 X_2$&  (1, 1, 0, 1)\\
     & $Z_1 X_2$ & (0, 0, 1, 0) \\
     & $X_1 X_2$ & (1, 0, 0, 1) \\
     & $Y_1 X_2$ & (0, 0, 1, 1) \\
     \hline
    \end{tabular}
    \label{tab:LUT_QEC_parallel}
\end{table*}

\begin{table*}[!tb]
    \centering
    \renewcommand*{\arraystretch}{1.7}
    \caption{\justifying \textbf{Lookup table for the fault-tolerant measurement of $X_\mathrm{L}^\mathrm{S} X_\mathrm{L}^\mathrm{T}$. } }
    \begin{tabular}{|c|c|}
    \hline
     flag error & resulting syndrome afterwards $(S_1, S_2, S_3, S_4)$\\
     \hline
     $X_2 Z_3$ & $ (1, 0, 1, 0)$ \\
     $Z_2 Z_3$ & $ (0, 1, 1, 1)$ \\
     $Y_2 Z_3$ & $ (1, 1, 1, 1)$ \\
     \hline
    \end{tabular}
    \label{tab:Flag_LUT_joint_X_measurement}
\end{table*}

\begin{table*}[!tb]
    \centering
    \renewcommand*{\arraystretch}{1.7}
    \caption{\justifying \textbf{Flag lookup table for the fault-tolerant measurement of $Z_\mathrm{L}^\mathrm{S}$. } }
    \begin{tabular}{|c|c|}
    \hline
     flag error & resulting syndrome afterwards $(S_1, S_2, S_3, S_4)$\\
     \hline
     $X_2 Z_4$ & $ (0, 0, 0, 1)$ \\
     $Z_2 Z_4$ & $ (1, 1, 0, 0)$ \\
     $Y_2 Z_4$ & $ (0, 1, 0, 0)$ \\
     \hline
    \end{tabular}
    \label{tab:Flag_LUT_Z_measurement}
\end{table*}

\section{Logical state preparation error characterization}\label{app:qst-characterization}
The observed noise PTM $R_\text{est}$ in a QPT experiment is affected by state preparation and measurement (SPAM) errors and differs from the operation PTM $R_\text{op}$, which we want to characterize
\begin{equation}
R_\text{est}=R_\text{meas}R_\text{op}R_\text{prep}.
\end{equation}
The SPAM errors are estimated by performing QPT on encoding and measuring a logical qubit, obtaining $R_\text{meas}R_\text{prep}$, but state preparation errors are in principle indistinguishable from measurement errors.

In the context of logical qubit tomography, we present a novel model-independent scheme to characterize the encoding circuit $R_\text{prep}$, provided that physical measurements errors are much smaller than circuit errors. These conditions are fulfilled in the presented setup since two-qubit gate errors are an order of magnitude larger than measurement errors as summarized in Tab.~\ref{tab:error_rates_simulation}. For CSS codes, we can leverage transversal destructive measurements together with readout error mitigation~\cite{bravyi2021mitigating} and simply assume $R_\text{meas}\approx I$ to obtain approximate values for $R_\text{prep}$. For non-CSS codes, we can replace the faulty logical measurement circuits by a single layer of physical measurements and perform physical quantum state tomography (QST). The output of QST is a density matrix and the expected values of logical operators are obtained by numerically evolving the matrix through the (ideal) logical measurement circuit. The expected values of logical operators are used to estimate a process matrix via logical QPT. Applying the same arguments as for the CSS case, we can assume that the physical measurement layer is reliable and identify the estimated process matrix with $R_\text{prep}$. For non-CSS codes, the number of different circuits required to characterize state preparation for $k$ logical qubits is thus $4^k3^N$, which is still unfeasible for experimental implementation.

\begin{table*}
    \centering
    \caption{\justifying \textbf{Postselected shot counts per reported fidelity}. Approximate total number of accepted shots per reported fidelity. If the number of accepted shots is below 1000, a range from lowest count to highest count across the respective measurement configurations (e.g. various initial states and readout bases) is given. Furthermore, the average acceptance rate is given and it is indicated if error correction (EC) or error detection (ED) was applied.}
    \begin{tabular}{|c|c|c|c|}
        \hline
         Experiment & Error handling & Acceptance rate & Accepted shots \\
         \hline
         Single-qubit SPAM (Fig.~\ref{fig:single_qubit_qec}(d)) & EC & 80 \% & $\approx 4 000$\\
         \hline
         Single-qubit quantum memory (Fig.~\ref{fig:single_qubit_qec}(e)) & EC & 75\% & $\approx 7 500$ \\
         \hline
         Two-qubit SPAM (Fig.~\ref{fig:results_crosstal}) & EC & 51\% & $\approx 5 100$ \\
         \hline
         Logical spectator error (Fig.~\ref{fig:results_crosstal}) & EC & 54\% & $\approx 1 080$ \\
         \hline
         Single-qubit SPAM (Fig.~\ref{fig:results_teleportation}(b)) & ED & 46\% & $\approx 2 300$ \\
         \hline
         Two-qubit quantum memory (Fig.~\ref{fig:results_teleportation}(c)) & ED & 0.85\% & 128-220\\
         \hline
         Teleportation (Fig.~\ref{fig:results_teleportation}(d)) & ED & 0.035\% & 18-25 \\
        \hline
        Magic state preparation (Fig.~\ref{fig:results_magic_states}(b)) & ED & 2.3\% & 186-259 \\
        \hline
        Entangling gate (Fig.~\ref{fig:results_magic_states}(c)) & ED & 0.36\% & 62-110 \\
        \hline
        Magic-state injection (Fig.~\ref{fig:results_magic_states}(d)) & ED & 0.023\% & 16-31 \\
        \hline
    \end{tabular}
    \label{tab:post-selected-total-shots}
\end{table*}

\clearpage
\bibliography{References}

\end{document}